\documentclass{aa}

\usepackage{graphicx}
\usepackage{caption}
\usepackage{subcaption}
\usepackage{natbib}
\usepackage{amsmath}
\usepackage{hyperref}
\usepackage{booktabs}

\begin{document}

\title{Creating retrogradely orbiting planets by prograde stellar fly-bys}

\author{Andreas Breslau\inst{\ref{mpifr}}
  \and Susanne Pfalzner\inst{\ref{mpifr}}\inst{\ref{fzj}}
}

\institute{Max-Planck-Institut f\"ur Radioastronomie, Auf dem H\"ugel 69, 53121 Bonn, Germany\label{mpifr}
  \and J\"ulich Supercomputing Center, Forschungszentrum J\"ulich, 52428 J\"ulich, Germany\label{fzj}\\
  \\
  \email{abreslau@mpifr.de}
}

\date{ }

\abstract{
  Several planets have been found that orbit their host star on retrograde orbits (spin-orbit angle $\varphi > 90^{\circ}$).
  Currently, the largest measured projected angle between the orbital angular momentum axis of a planet and the rotation axis of its host star
  has been found for \mbox{HAT-P-14b} to be $\approx 171^{\circ}$.
  One possible mechanism for the formation of such misalignments is through long-term interactions between the planet and other planetary or stellar companions.
  However, with this process, it has been found to be difficult to achieve retrogradely orbiting planets, especially
  planets that almost exactly counter-orbit their host star ($\varphi \approx 180^{\circ}$) such as HAT-P-14b.
  By contrast, orbital misalignment can be produced efficiently by perturbations of planetary systems that are passed by stars.
  Here we demonstrate that not only retrograde fly-bys, but surprisingly, even {\bf prograde} fly-bys can induce retrograde orbits.
  Our simulations show that depending on the mass ratio of the involved stars,
  there are significant ranges of planetary pre-encounter parameters for which counter-orbiting planets are the natural consequence.
  We find that the highest probability to produce counter-orbiting planets ($\approx 20\%$) is achieved with close prograde, coplanar fly-bys of an equal-mass perturber
  with a pericentre distance of one-third of the initial orbital radius of the planet.
  For fly-bys where the pericentre distance equals the initial orbital radius of the planet,
  we still find a probability to produce retrograde planets of $\approx 10$~\%
  for high-mass perturbers on inclined ($60^{\circ} < i < 120^{\circ}$) orbits.
  As usually more distant fly-bys are more common in star clusters, 
  this means that inclined fly-bys probably lead to more retrograde planets than those with inclinations $< 60^{\circ}$.
  Such close fly-bys are in general relatively rare in most types of stellar clusters,
  and only in very dense clusters will this mechanism play a significant role.
  The total production rate of retrograde planets depends then on the cluster environment.
  Finally, we briefly discuss the application of our results to the retrograde minor bodies in the solar system and to the formation of retrograde moons during the planet-planet scattering phase.
}


\maketitle

\section{Introduction}
\label{sec:intro}

Since the discovery of the first exoplanets in the 1990s \citep[e.g.][]{1995Natur.378..355M}, more than three thousand exoplanets have been confirmed.\footnote{Researches
  for exoplanets and their orbital properties for this paper have been conducted using  the services of exoplanet.eu \citep{2011A&A...532A..79S} and exoplanets.org \citep{2014PASP..126..827H}.}
Many of these planets differ considerably in their properties (e.g. mass, temperature, and orbital period) from those in the solar system.

One interesting property is the relative alignment or misalignment between the orbital angular momentum vector of an exoplanet
and the rotation axis of its stellar host,
because it allows the inference of information about the formation and subsequent evolution of the planetary system.
One method to measure this misalignment
was originally developed by \citeauthor{1924ApJ....60...15R} and \citeauthor{1924ApJ....60...22M} in 1924 for eclipsing binary stars.
In the application to exoplanet research, the
Rossiter-McLaughlin effect exploits the
occultation of part of the rotating stellar surface by a transiting planet.
Since the light from the rotating stellar surface is Doppler shifted, a shadowing of a fraction of that light results in a measurable variation of the radial velocity
obtained from the Doppler-shifted light of the star.
Together with other parameters of star and planet, this information can be used to determine the projected angle $\lambda$ between the planetary angular momentum vector and the stellar rotation axis
\cite[e.g.][]{2000A&A...359L..13Q}.

\begin{table*}[ht]
  \caption{Exoplanets with measured projected alignment angle relative to the stellar rotation axis of $\lambda > 90^{\circ}$.}
  \label{tab:retrograde_planets}
  \centering
  \begin{tabular}{lllll}
    \toprule
    Name             & $\lambda$ [$^{\circ}$] & Ref  & $a$ [au]                     & $e$ \\
    \midrule
    WASP-79 b         & $-\hphantom{0}99.1~(+4.1,-3.9)$       & [1]  & $0.0539 \pm 0.0009$          & $0.0$ \\
    \, \, \, ''       & $-\hphantom{0}95.2~(+0.9,-1.0)$       & [2]  &                              & \\
    WASP-121 b        & $\hphantom{-}102.2~(+5.5,-5.3)$       & [3]  & $0.0254 \pm 0.0005$          & $0.0$ \\
    HAT-P-11 b        & $\hphantom{-}103~(+23,-19)$           & [4]  & $0.0530 \pm 0.0008$          & $0.265~(+0.0003,-0.0009)$ \\
    Kepler-63 b       & $-115$                                & [5]  & $0.08\phantom{00} \pm 0.002$ & -- \\
    KELT-17 b         & $-115.9 \pm 4.1$                      & [6]  & $0.0488 \pm 0.0006$          & -- \\
    HAT-P-18 b        & $\hphantom{-}132 \pm 15$              & [7]  & $0.0559 \pm 0.0007$          & $0.106~(-0.084,+0.15)$ \\
    HAT-P-7 b         & $-132.6~(+10.5,-16.3)$                & [8]  & $0.0379 \pm 0.0004$          & $0.0$ \\
    WASP-15 b         & $\hphantom{-}139.6~(+5.2,-4.3)$       & [9]  & $0.0499 \pm 0.0018$          & $0.0$ \\
    WASP-8 b          & $-143.0~(+1.6,-1.5)$                  & [10] & $0.0801 \pm 0.0016$          & $0.31 \pm 0.002$ \\ 
    WASP-17 b         & $-150$                                & [11] & $0.0515 \pm 0.0003$          & $0.028~(+0.015,-0.018)$ \\
    \, \, \, ''       & $\hphantom{-}148.5~(+5.1,-4.2)$       & [9]  &                              & \\
    \, \, \, ''       & $\hphantom{-}167.4 \pm 11.2$          & [12] &                              & \\
    WASP-94 A b       & $\hphantom{-}151 \pm 20$              & [13] & $0.055\phantom{0} \pm 0.001$ & $0.0$ \\ 
    WASP-2 b          & $\hphantom{-}153~(+11,-15)$           & [9]  & $0.031\phantom{0} \pm 0.011$ & $0.0$ \\
    WASP-167 b        & $-165  \pm 5$                         & [14] & $0.0365 \pm 0.0006$          & -- \\
    HAT-P-6 b         & $\hphantom{-}166 \pm 10$              & [15] & $0.0524 \pm 0.0009$          & $0.0$ \\
    HAT-P-14 b        & $-170.9 \pm 5.1$                      & [16] & $0.0594 \pm 0.0004$          & $0.095 \pm 0.011$ \\
    \bottomrule
  \end{tabular}
  \tablefoot{The table is sorted to ascending absolute values of $\lambda$.
    The values for semi-major axes $a$ and eccentricities $e$ are taken from exoplanet.eu.}
  \tablebib{[1]~\citet{2017AJ....154..137J}; [2]~\citet{2017MNRAS.464..810B}; [3]~\citet{2016MNRAS.458.4025D};
    [4]~\citet{2011PASJ...63S.531H}; [5]~\citet{2013ApJ...775...54S}; [6]~\citet{2016AJ....152..136Z};
    [7]~\citet{ 2014A&A...564L..13E}; [8]~\citet{2009PASJ...61L..35N}; [9]~\citet{2010A&A...524A..25T};
    [10]~\citet{2017A&A...599A..33B}; [11]~\citet{2010ApJ...709..159A}; [12]~\citet{2010ApJ...722L.224B};
    [13]~\citet{2014A&A...572A..49N}; [14]~\citet{2017MNRAS.471.2743T}; [15]~\citet{2011A&A...527L..11H};
    [16]~\citet{2011AJ....141...63W} }
\end{table*}

Alternatively, the relative angle between the stellar spin axis and the orbital angular momentum axis can be obtained
by exploiting effects of
star spot occultations \citep{2011ApJ...740L..10N,2013A&A...549A..35O} or gravity darkening
\mbox{\citep{2011ApJS..197...10B, 2014ApJ...786..131A}}.
These methods
make use of the extensive photometric data bases of the {\em Kepler} and {\em CoRoT} (Convection, Rotation et Transits planétaires) missions.
For the star spot method, the long-term light curve is used to fit a model of a rotating star with spots to which in turn the transit light curves are fitted.
The gravity-darkening method relies on the fact that the stellar surface brightness of fast-rotating stars is higher at the stellar poles than near the equator.
This results in characteristic shapes for the transit light curves depending on the spin-orbit alignment.
With these methods, not only the projected angle $\lambda$, but the true spin-orbit angle $\varphi$ can be obtained.\footnote{In the literature,
  different conventions are used for the spin-orbit angle.
  Sometimes it is denoted with $\psi$, while in other cases, $\psi$ is used for the stellar obliquity. 
  To avoid confusion with the stellar obliquity, and because we use $\psi$ for another quantity,
  we denote the true spin-orbit angle with $\varphi$, as did \citet{2011ApJS..197...10B} and \citet{2014ApJ...786..131A}.
  Since we are here interested in the angle enclosed by these two axes, and not in
  the absolute orientation of particle orbit and stellar spin axis in space,
  we consider the smallest angle between the two axes, neglecting any sense of rotation.
  Therefore, our spin-orbit angles are in the range $0^{\circ} \le \varphi \le 180^{\circ}$.
}

At the time of writing, 
considerable spin-orbit misalignment has been measured for approximately 100 exoplanets
\citep[e.g.][]{2009ApJ...696.1230F, 2009PASP..121.1104J, 2010A&A...524A..25T, 2011A&A...533A.113M, 2012ApJ...744..189A, 2015A&A...579A.136M}.
About 15 of these have been found with $\lambda > 90^{\circ}$ (listed in Table~\ref{tab:retrograde_planets}).
All planets for which $\lambda$ has been measured are transiting in front of their host star, that is, their orbits are seen edge-on.
The true spin-orbit angle $\varphi$ for $0^{\circ} < \lambda < 90^{\circ}$ is therefore usually also in the range $0^{\circ}$ to $90^{\circ}$, but may be larger than $\lambda$,
for $90^{\circ} < \lambda < 180^{\circ}$, it is usually also in the range $90^{\circ}$ to $180^{\circ}$, but may be smaller \citep[e.g.][]{2014A&A...567A..42C}.
The probability is therefore high that for these 15 planets with $\lambda > 90^{\circ}$ also $\varphi > 90^{\circ}$,
which means that they move on retrograde orbits.

These large misalignments disagree with the assumed formation process for planetary systems,
according to which the star and planets form from the same rotating disc of gas and remain unaltered thereafter.
This means that either the spin axis of the stars or the orbital angular momenta of the planets have to be changed by some process,
either during the formation phase or afterwards.

Misalignment between planetary orbits and stellar rotation axes can already be
caused in the planet formation phase by
a tilting of the protoplanetary discs relative to the stars.
Such a tilting could be caused by long-term interactions with a binary companion \citet{2012Natur.491..418B},
close fly-bys of other stars \mbox{\citep{2010MNRAS.401.1505B}},
non-coplanar capture of gas to the disc \mbox{\citep{2011MNRAS.417.1817T}},
or even a tilting of the stellar rotation axis relative to the disc as a result of an interaction of the stellar magnetic field with the disc \mbox{\citep{2011MNRAS.412.2790L}}.
Planets forming from these misaligned discs would then also be misaligned.

When the planets have already formed and the protoplanetary discs no longer exist, the planetary orbital angular momenta may be altered directly by other processes such as
gravitational scattering between the planets \citep{2008ApJ...686..580C}
or long-term perturbations by a stellar or planetary companion in the form of Kozai-Lidov migration \citep[e.g.][]{2007ApJ...669.1298F}.
Although \citet{2014ApJ...785..116L} have shown that even the production of counter-orbiting\footnote{Similar to
  \citet{2016ApJ...820...55X}, we use here the term 'counter-orbiting' for those planets with $\varphi \approx 180^{\circ}$
  to distinguish them from 'retrograde' planets with $\varphi > 90^{\circ}$.
  For practical reasons, we denote all orbits with $170^{\circ} \le \varphi \le 180^{\circ}$ as 'counter-orbiting'.}
planets is possible,
the production of retrogradely orbiting planets has been found to be very difficult in general for these processes
\citep[e.g.][]{2011Natur.473..187N,2016ApJ...820...55X}.

Another process altering the spin-orbit alignment when the planets have already formed is the perturbation of planetary systems by stellar fly-bys in star clusters
\citep[e.g.][]{2011MNRAS.411..859M, 2013MNRAS.433..867H}.
\citeauthor{2013MNRAS.433..867H} found in their investigations that at the end of the simulations, a small fraction of planets moved on retrograde orbits.
In the particular case they investigated, $< 5\%$ of the planets were finally on retrograde orbits around their host star.

Similar to \citet{2013MNRAS.433..867H}, we investigate here the production of retrograde planetary orbits by the perturbation of a planet
that is initially and finally bound to its host star by the fly-by of another star.
While \citet{2013MNRAS.433..867H} have shown that the production of retrograde orbits by stellar fly-bys is in principle possible,
we explicitly specify the conditions leading to counter-orbiting planets such as \mbox{HAT-P-14b}.
Additionally, we determine the occurrence rate of counter-orbiting planets depending on the fly-by parameters and the initial conditions of the perturbed bodies.

In Sect.~\ref{sec:method} the numerical method is explained and a special reference frame is introduced that is used throughout the paper to depict perturbed particle trajectories.
In Sect.~\ref{sec:results} the trajectories leading to particles that counter-orbit their host are explained for some sample mass-ratios.
Then the probabilities for these trajectories are analysed first for coplanar fly-bys and then averaged over the inclinations of the perturber orbit.
In Sect.~\ref{sec:discussion} the limitations of the method are discussed.
Finally, in Sect.~\ref{sec:application} the potential applicability of the results to the retrograde bodies in the solar system is considered.

\section{Method}
\label{sec:method}

We performed three-dimensional, purely gravitational, numerical three-body simulations of
interactions between mass-less tracer particles and two massive objects.
The tracer particles initially orbit on circular Keplerian orbits around
an object with mass $M_{\mathrm{h}} $.
Here and in the following,
variables with the index 'p' denote properties of the perturber or its orbit,
while variables with index 'h' denote properties of the host, and variables without index usually refer to the particle.

This initial configuration is perturbed by the fly-by of another object with mass $M_{\mathrm{p}}  = m \cdot M_{\mathrm{h}} $ on a parabolic orbit
with pericentre distance of $r_{\mathrm{p,peri}}  = 1$.
The initial positions of the mass-less particles are given in virtual pericentre positions (VPP),
which are the positions where the particles would be without the perturbation
at the moment of the pericentre passage of the perturber.
This is analogous to the well-known impact parameter for scattering processes
that defines an initial condition through the distance at which the particle would pass the target if it were not influenced.
For a more detailed description of the VPP space, see \citet{2017A&A...599A..91B}.\footnote{To improve readability,
  we correct here a small inconsistency in \citet{2017A&A...599A..91B}:
  in the method, all lengths in that paper were scaled with $r_{\mathrm{p,peri}} $.
  Later on, for example $r_{\mathrm{vpp}} /r_{\mathrm{p,peri}} $ was used even though $r_{\mathrm{vpp}} $ would have been sufficient.
  Here, we omit the $/r_{\mathrm{p,peri}} $ where possible.
}

The initial distance between the host and the perturber, $r_{\mathrm{p,init}} $, is determined by defining a
maximum relative initial force influence of the perturber on the particle compared to the force from the host of
\begin{align}
  F_{\mathrm{p}}/F_{\mathrm{h}} < \epsilon\label{eq:force}.
\end{align}
Here, the value $\epsilon = 10^{-4}$ was chosen.
For a given orbital eccentricity, $e_{\mathrm{p}}$, the true anomaly $\theta_{\mathrm{p,init}} $ can be found from $r_{\mathrm{p,init}} $ using the relation
\begin{align}
  r_{\mathrm{p,init}}  = \frac{r_{\mathrm{p,peri}}  (1 + e_{\mathrm{p}})}{1+ e_{\mathrm{p}} \cos(\theta_{\mathrm{p,init}} )}.
\end{align}
Because of the parabolic orbit of the perturber, here $e_{\mathrm{p}} = 1 $.
The time of flight from the initial position of the perturber to pericentre is given by
\begin{align}
  t_{\mathrm{peri}}  - t_{\mathrm{init}}  = \sqrt{\frac{ 2 r_{\mathrm{p,peri}} ^3 }{ \mu } } \left(E - e_{\mathrm{p}} \sin{E} \right)\label{eq:time_of_flight},
\end{align}
with the standard gravitational parameter for the two-body problem $\mu = G (M_{\mathrm{h}}  + M_{\mathrm{p}} )$ and the eccentric anomaly $E$.

The initial position of the particle follows then from the sampled VPP ($x_{\mathrm{vpp}} $ and $y_{\mathrm{vpp}} $ or $r_{\mathrm{vpp}} $ and $\theta_{\mathrm{vpp}} $, respectively) according to
\begin{align}
  \begin{split}
    \theta_{\mathrm{init}}  &= \left[ \theta_{\mathrm{vpp}}  - \omega_{\mathrm{0}}  \, (t_{\mathrm{peri}}  - t_{\mathrm{init}} ) \right]\mod 2 \pi\label{eq:init_pos_cyl}\\
    r_{\mathrm{init}}  &= r_{\mathrm{vpp}} ,
  \end{split}
\end{align}
with $\omega_{\mathrm{0}}  = \sqrt{G M_{\mathrm{h}}  / r_{\mathrm{vpp}} ^3}$ being the Keplerian velocity at $r_{\mathrm{vpp}} $ around the host.

The trajectory of each particle was integrated individually with the adaptive LSODA integrator from the ODEPACK library \citep{ODEPACK}.
The trajectory integration was ended when the particle had settled into a stable orbit around host or perturber,
or departed from the centre of mass on a stable hyperbolic orbit.
An orbit was defined as stable when the orbital elements did not change more than $1\%$ for $\delta t = (t_{\mathrm{peri}} -t_{\mathrm{init}} ) / 50$.
A more detailed description can be found in \citet{2017A&A...599A..91B}.

Our simulations are pure gravitational interactions of point-like particles,
thus the host star has no internal angular momentum and therefore no spin axis.
We define the low-mass particle initial orbital angular momentum axis (the z-axis) and the host spin axis to be initially aligned.
The final spin-orbit angle $\varphi$ is then determined by measuring the final angle between the particle orbital angular momentum and this axis.

\subsection{Timescale}

For convenience, we define $t = 0$ as the moment of the perturber pericentre passage.
As a reference duration, we use the time needed for the perturber to move $90^{\circ}$ along its orbit after pericentre passage.
The distance $d$ between host and perturber $90^{\circ}$ {\bf after} pericentre passage 
is $ d = 2\,r_{\mathrm{p,peri}} $ for a parabolic orbit.
Because this point (at (0,\,2)) is known as semilatus rectum, we denote it as $SLR$,
the point $90^{\circ}$ {\bf before} pericentre passage (at (0,\,-2)) as $-SLR$.
The time that the perturber needs from pericentre to $SLR$, we denote with $t_{\mathrm{slr}}$, which is given by
\begin{align}
  t_{\mathrm{slr}} = 4/3 \sqrt{\frac{2 r_{\mathrm{p,peri}} ^3}{ \mu} }. \label{eq:tslr}
\end{align}

\subsection{Depiction of particle trajectories}
\label{sec:rotating_frame}

In this work, our aim is among other things to show trajectories of perturbed particles.
The widely used method of plotting the coordinates of particle, host, and perturber in normal space has
two disadvantages:
On the one hand, it is difficult to represent the temporal development of the involved particles.
Even when we plot, for example, in the host centred reference frame, the positions of two objects have to be shown for every time step.
For an average perturbed particle trajectory, these positions have to be shown for several time steps,
either with one image showing two coordinates for every time step or by combining them into one image showing two trajectories.
These trajectories usually cross each other.
Another difficulty arises from the huge difference in scale throughout the perturbation process.
The factor between the full picture and the smallest detail of the interaction may span several orders of magnitude.

Here we use instead a rotating and rescaling reference frame (R3F) \citep[see e.g.][]{1975CeMec..12..175H},
which makes it possible to show the entire fly-by and circumvents the above-described difficulties.
For each time step, first the coordinates of host, perturber, and particle are scaled such that the distance between host and perturber equals \mbox{one ($1$)}.
Then the whole frame is shifted and rotated until the host is in (0,\,0) and the perturber in (1,\,0).
The resulting Cartesian coordinates we denote with $\chi$ and $\Psi$.
See Sect.~\ref{sec:more_rotating} in the Appendix for a brief definition or, for example,
\citet{Alvarez2006} for a more formal derivation.

After the transformation to this reference frame, only one time-dependent coordinate remains, illustrating
the movement of the particle. Initially, the particle moves very closely around the host,
the scaled initial orbital radius is $< 1/30$ for $m = 0.1$, for instance.
Only relatively shortly before pericentre passage of the perturber may the particle leave the region with $r = 1$ around the host.
Because of the automatic rescaling of the R3F,
the interactions of the particle with host and perturber close to the pericentre passage of the perturber
are usually as well visible as the asymptotic settling of the particle either into a bound orbit around host or perturber,
or into an unbound orbit.

\begin{figure}[t!]
  \centering
   \includegraphics[width=0.9\hsize]{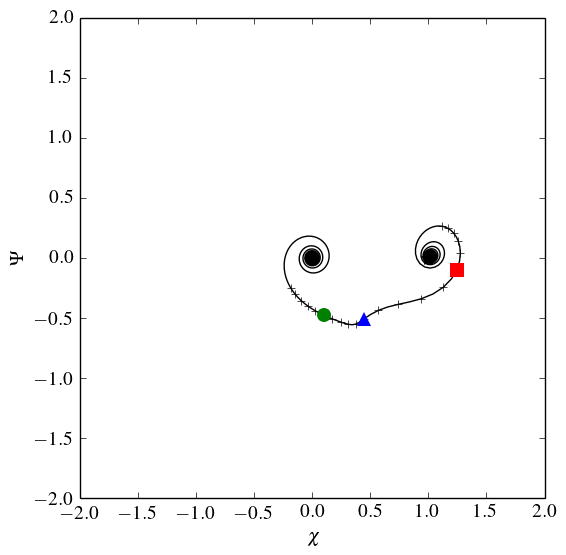}
   \caption{Interaction where the particle is captured by the perturber in a prograde, coplanar, parabolic fly-by in the R3F.
     The green circle, blue triangle, and red square mark the positions of the particle for $t = -t_{\mathrm{slr}}$, $t= 0$, and $t = t_{\mathrm{slr}}$ respectively.
     The definition of the Cartesian coordinates of the R3F, $\chi$ and $\Psi$,
     can be found in Sect.~\ref{sec:more_rotating} in the Appendix.
   }
   \label{fig:rotating_sample}
\end{figure}

\begin{figure}[ht!]
  \centering
  \begin{minipage}[t]{\hsize}
    \vspace{0pt}
    \begin{subfigure}[t]{\textwidth}
      \begin{minipage}[t]{0.04\textwidth}
        \vspace{0pt}
          \caption{}
          \label{fig:retrograde_vs_inclination}
      \end{minipage}
      \hfill
      \begin{minipage}[t]{0.95\textwidth}
        \vspace{0pt}
        \includegraphics[width=\hsize]{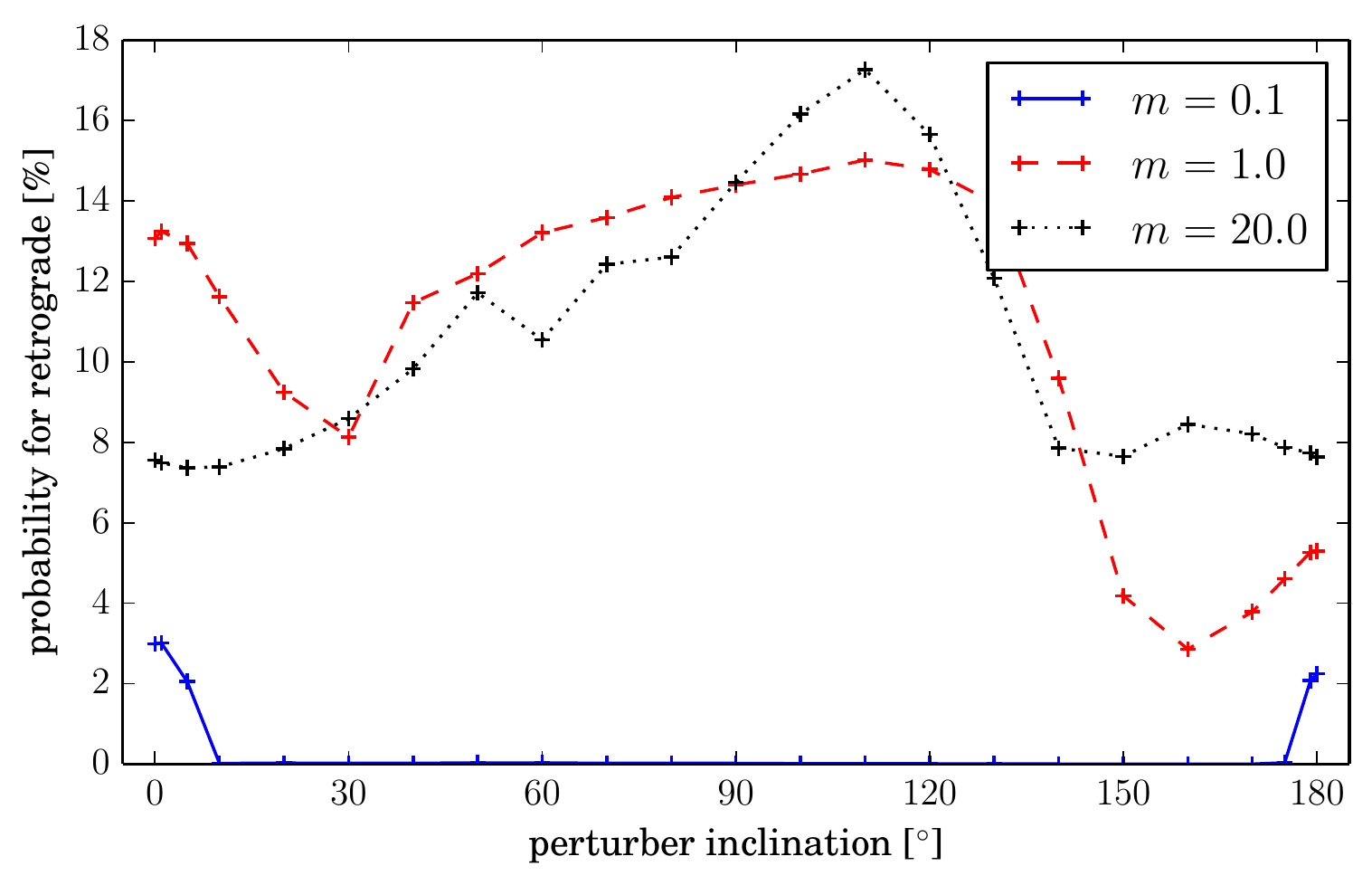}
        \end{minipage}
    \end{subfigure}
  \end{minipage}
  \begin{minipage}[t]{\hsize}
    \vspace{0pt}
    \begin{subfigure}[t]{\textwidth}
      \begin{minipage}[t]{0.04\textwidth}
        \vspace{0pt}
        \caption{}
        \label{fig:RC_vs_inclination}
      \end{minipage}
      \hfill
      \begin{minipage}[t]{0.95\textwidth}
        \vspace{0pt}
        \includegraphics[width=\hsize]{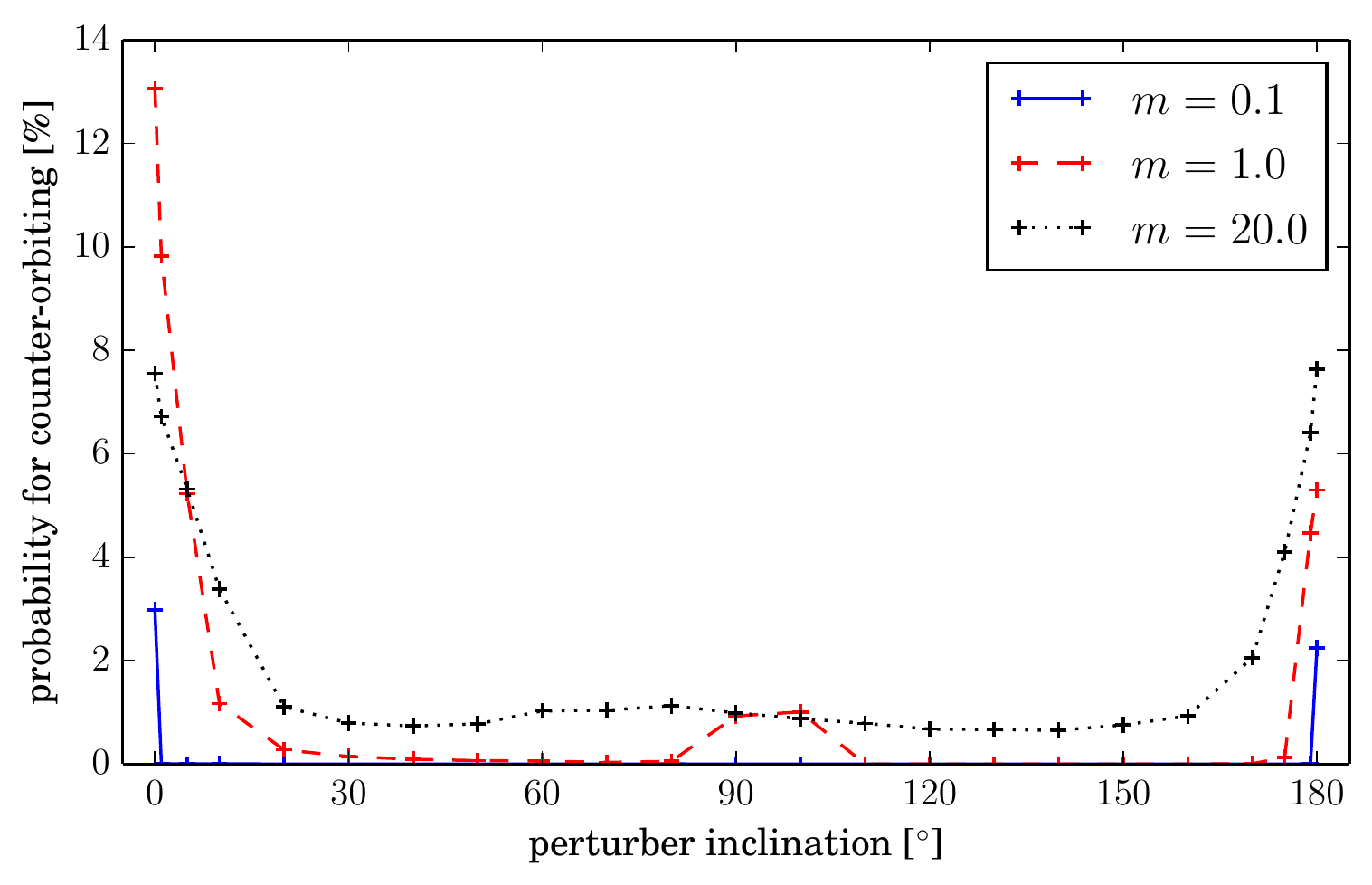}
      \end{minipage}
    \end{subfigure}
  \end{minipage}
  \caption{Probability for a planet (or disc material) to become\\
    {\bf a)} retrograde ($\varphi > 90 ^{\circ}$) and\\
    {\bf b)} counter-orbiting ($\varphi > 170 ^{\circ}$)
    depending on the inclination of the perturber orbit and the perturber mass.
  }
  \label{fig:probs_vs_mass}
\end{figure}

Figure~\ref{fig:rotating_sample} shows an exemplary interaction where the particle is captured by the perturber
in the R3F.
Showing the capture of a particle by the perturber is especially difficult in normal space
because in general, neither the host-centred frame nor the perturber-centred frame alone can be used to show the full trajectory.
The blue triangle marks the position of the particle when the perturber is in pericentre ($t= 0$),
the green circle and red square show the positions of the particle for $t = -t_{\mathrm{slr}}$ and $t = t_{\mathrm{slr}}$, respectively.
The crosses mark the particle positions from $t = -2\, t_{\mathrm{slr}}$ to $t = 2\, t_{\mathrm{slr}}$ in steps of $dt = 0.2\, t_{\mathrm{slr}}$.

In the case shown here,
the particle moves counter-clockwise around the host while the perturber approaches (spiral movement around the host).
When the perturber reaches pericentre at $t= 0$ (particle position at that time indicated by the blue triangle),
the particle approaches the region that is gravitationally dominated by the perturber.
The particle enters the gravitational field of the perturber and starts orbiting the perturber counter clockwise
while the perturber departs.

For a given mass-ratio between host and perturber and eccentricity of the perturber orbit,
the position of the perturber can be determined for every time $t$.
With this information, the particle position in the R3F (given by $\chi$ and $\Psi$)
can be transformed back into the time-dependent coordinates in normal space of all three objects.

\section{Results}
\label{sec:results}

Even though \citet{2013MNRAS.433..867H} have shown that the production of retrograde planetary orbits through fly-bys is in general possible,
the total number of retrograde planets they found in their study (i.e. $5$) is
far too low to obtain reliable statistics concerning the general probability and parameter dependence of such processes.
Therefore, we first examine the effect of a fly-by on a population of planets.

\subsection{Total probabilities for retrograde orbits from fly-bys}
\label{sec:total_probabilities}

Fly-bys that may change the orbits of planets only occur with a significant probability in
young star clusters where the stellar density is sufficiently high
(age up to $\approx 10$~Mry, see also discussion in Sect.~\ref{sec:discussion}).
Since the exoplanets observed so far are, in contrast, relatively old (from hundreds of~Myr up to several~Gyr),
models derived from the observed orbital radius distribution \citep[e.g.][]{2016ApJ...819..125C}
are not suitable for a direct comparison.
The synthesis of planet populations, on the other hand, is a very complex process depending on a large number of parameters \citep[see e.g.][and references therein]{2016SSRv..205...77B}.
This field is still in a very early stage, and ready-to-use algorithms for the creation of populations a few Myr old are not yet available.

Therefore, we naively assume in the following that the initial orbital radii of planets are equally distributed.
This distribution is motivated by the $1/r$ mass surface density distribution, which is often considered a good compromise for simulations of protoplanetary discs
\citep[e.g.][and references therein]{2012A&A...538A..10S}.
The $1/r$ mass surface density distribution corresponds to equal masses per radial bin.

We assume a population of planets with orbital radii between $1$~au and $100$~au
that is perturbed by a fly-by of a star on a parabolic orbit with pericentre distance of $10$~au.
This translates into initial radii as defined in Sect.~\ref{sec:method} of $0.1 < r_{\mathrm{init}}  < 10$.
To use the full information of our simulation data, we place one planet in each bin of our data grid.
This is spanned by 180 angular and 228 radial bins, resulting in a population of 41040 planets.
All planets are initially on circular orbits.
For the orbit of the perturber, we consider inclinations from $0^{\circ}$ to $180^{\circ}$ in steps of $10^{\circ}$.
To improve the resolution for almost coplanar perturbations, we also include inclinations of $1^{\circ}, 5^{\circ}, 175^{\circ}$, and $ 179^{\circ}$.
Inclinations of the orbit of the perturber are throughout measured relative to the plane of the orbiting planet (the x-y-plane).
The ascending node always lies on the positive \mbox{x-axis}.
Because of the otherwise much larger size of the parameter space, we consider here only an argument of periapsis of $\omega = 0$.
Thus, the pericentre of the perturber orbit is always on the positive x-axis.

\begin{table}[t!]
  \centering
  \begin{tabular}{lll}
    \toprule
    $m$      & $i = 0^{\circ}$     & $i = 180^{\circ}$ \\
    \midrule
    $\hphantom{0}0.1$ & $\hphantom{1}3$~\%  & $2$~\% \\
    $\hphantom{0}1.0$ & $13$~\%             & $5$~\% \\
    $20.0$            & $\hphantom{1}8$~\%  & $8$~\% \\
    \bottomrule
  \end{tabular}
  \caption{Probability for a particle with $0.1 < r_{\mathrm{init}}  < 10$ to become counter-orbiting depending on mass ratio and inclination of the perturber orbit.}
  \label{tab:RC_probability_total}
\end{table}

Figure~\ref{fig:retrograde_vs_inclination} shows the total probability for a planet from our sample population
to end up on a retrograde orbit ($\varphi > 90^{\circ}$) around its host depending on the inclination of the perturber orbit for three representative sample mass-ratios:
$m = 0.1$ as an example for a low-mass perturber,
$m = 20.0$ for a high-mass perturbation,
and the equal-mass case with $m = 1.0$.
We emphasise that our spin-orbit angles are in the range $0^{\circ} \le \varphi \le 180^{\circ}$.
It can be seen that no simple trend exists between the three mass-ratios.
The probability for $m = 0.1$ is significantly non-zero only for perturber orbits with inclinations $i < 10^{\circ}$ and $i > 175^{\circ}$.
For $m = 20.0$, the probability is $\approx 8$~\% for $i \approx 0^{\circ}$ and $i \approx 180^{\circ}$
and has a maximum in between of $\approx 17$~\% for $i \approx 110^{\circ}$.
For $m = 1.0$, the probability is much more complicated.
It starts at $\approx 13$~\% for $i \approx 0^{\circ}$, decreases then to $\approx 8$~\% for $i \approx 30^{\circ}$,
rises again to $\approx 13 \text{--} 15$~\% for $60^{\circ} \le i \le 130^{\circ}$, and falls to an absolute minimum of $\approx 3$~\% for $i \approx 160^{\circ}$
before it rises again to $\approx 5$~\% for $i \approx 180^{\circ}$.

We note that for the two coplanar cases, $i = 0^{\circ}$ and $i = 180^{\circ}$, the problem is just two-dimensional,
while it is three-dimensional for all other inclinations.
The excitation of misalignment between host star and planet in the three-dimensional case is not very surprising.
Since the perturber does not move in the same plane as the particle, it may add some momentum to the z-velocity component of the particle.
This causes the final orbital plane of the particle to differ from the initial one.
As also the original x- and y-components of the particle velocity are affected, the final particle orbit may be retrograde.

In the coplanar cases, in contrast, the excitation of retrograde orbits {\bf must} occur in the plane, because none of the involved particles has any z-velocity component.
As a consequence, in these cases, the retrograde particles do not just have $\varphi > 90^{\circ}$ but $\varphi = 180^{\circ}$ and are thus counter-orbiting.
Figure~\ref{fig:RC_vs_inclination} shows the probabilities to produce counter-orbiting ($\varphi \ge 170^{\circ}$) particles depending on the mass ratio and the perturber inclination.
It can be seen that the probability is very low ($< 2\%$)
for perturber orbits with inclinations $10^{\circ} \le i \le 170^{\circ}$ for all three sample mass-ratios.
Only close to $i = 0^{\circ}$ and $i = 180^{\circ}$
are the probabilities significantly higher.
The approximate values for $i = 0^{\circ}$ and $i = 180^{\circ}$ for all three mass ratios are summarised in Table~\ref{tab:RC_probability_total}.

The total probabilities to form retrograde ($\varphi > 90 ^{\circ}$) or counter-orbiting ($\varphi > 170 ^{\circ}$) particles
in prograde ($i \le 90^{\circ}$) and retrograde ($i \ge 90^{\circ}$) fly-bys are listed in Table~\ref{tab:probs_for_prograde_retrograde_encounter}.
The values were obtained by integrating the data illustrated in Figs.~\ref{fig:retrograde_vs_inclination} and \ref{fig:RC_vs_inclination} with the trapezoidal rule.
The only case where the probability of the creation of a retrograde orbit is higher for a retrograde fly-by than for a prograde fly-by is for a high-mass perturber ($m = 20.0$).
In all other cases the probabilities are higher for prograde fly-bys.

\begin{table}[t!]
  \centering
  \begin{tabular}{ccccc}
    \toprule
    $m$      & \multicolumn{2}{c}{retrograde}                  & \multicolumn{2}{c}{counter-orbiting} \\
                      & $i \le 90^{\circ}$     & $i \ge 90^{\circ}$     & $i \le 90^{\circ}$ & $i \ge 90^{\circ}$ \\
    \midrule
    $\hphantom{0}0.1$ & $\hphantom{1}0.22$~\%  & $\hphantom{1}0.08$~\%  & $0.02$~\%          & $0.01$~\% \\
    $\hphantom{0}1.0$ & $11.96$~\%             & $\hphantom{1}9.86$~\%  & $0.84$~\%          & $0.32$~\% \\
    $20.0$            & $10.21$~\%             & $11.60$~\%             & $1.57$~\%          & $1.25$~\% \\
    \bottomrule
  \end{tabular}
  \caption{Probability to form retrograde ($\varphi > 90 ^{\circ}$) or counter-orbiting ($\varphi > 170 ^{\circ}$) particles
    for prograde ($i \le 90^{\circ}$) and retrograde ($i \ge 90^{\circ}$) fly-bys depending on the mass-ratio.}
  \label{tab:probs_for_prograde_retrograde_encounter}
\end{table}

\begin{figure}[ht!]
  \centering
  \begin{minipage}[t]{\hsize}
    \vspace{0pt}
    \begin{subfigure}[t]{\textwidth}
      \begin{minipage}[t]{0.04\textwidth}
        \vspace{0pt}
          \caption{}\label{fig:map_0.1}
      \end{minipage}
      \hfill
      \begin{minipage}[t]{0.9\textwidth}
        \vspace{0pt}
        \includegraphics[width=\textwidth]{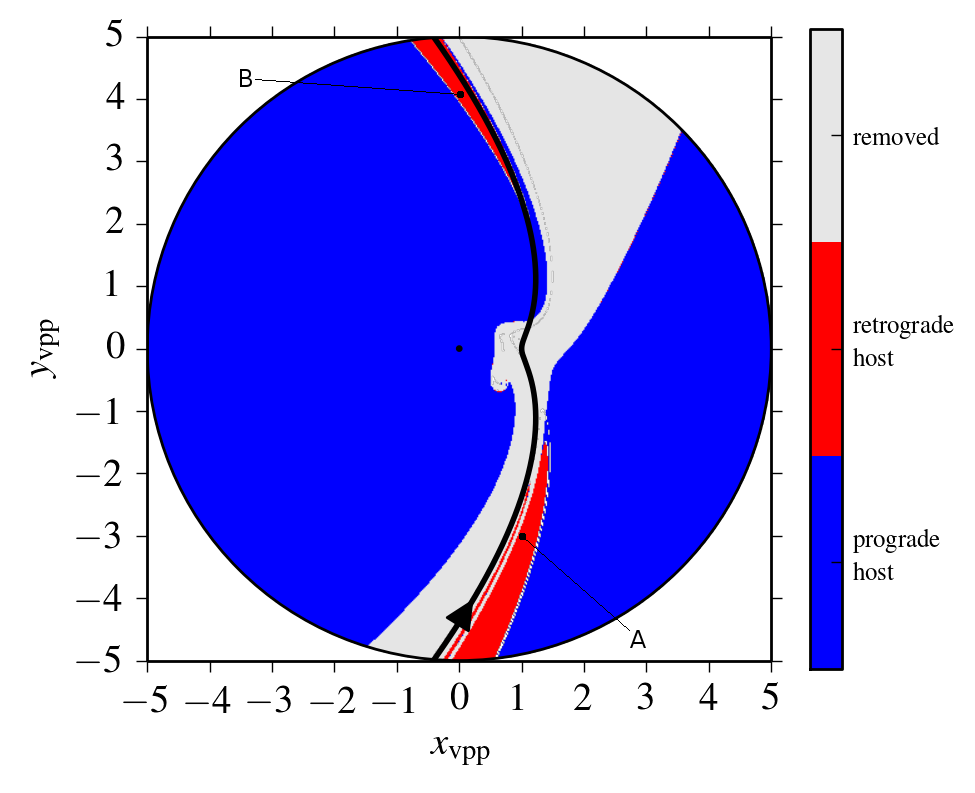}
        \end{minipage}
    \end{subfigure}
  \end{minipage}
  \begin{minipage}[t]{\hsize}
    \vspace{0pt}
    \begin{subfigure}[t]{\textwidth}
      \begin{minipage}[t]{0.04\textwidth}
        \vspace{0pt}
        \caption{}\label{fig:map_1}
      \end{minipage}
      \hfill
      \begin{minipage}[t]{0.9\textwidth}
        \vspace{0pt}
        \includegraphics[width=\textwidth]{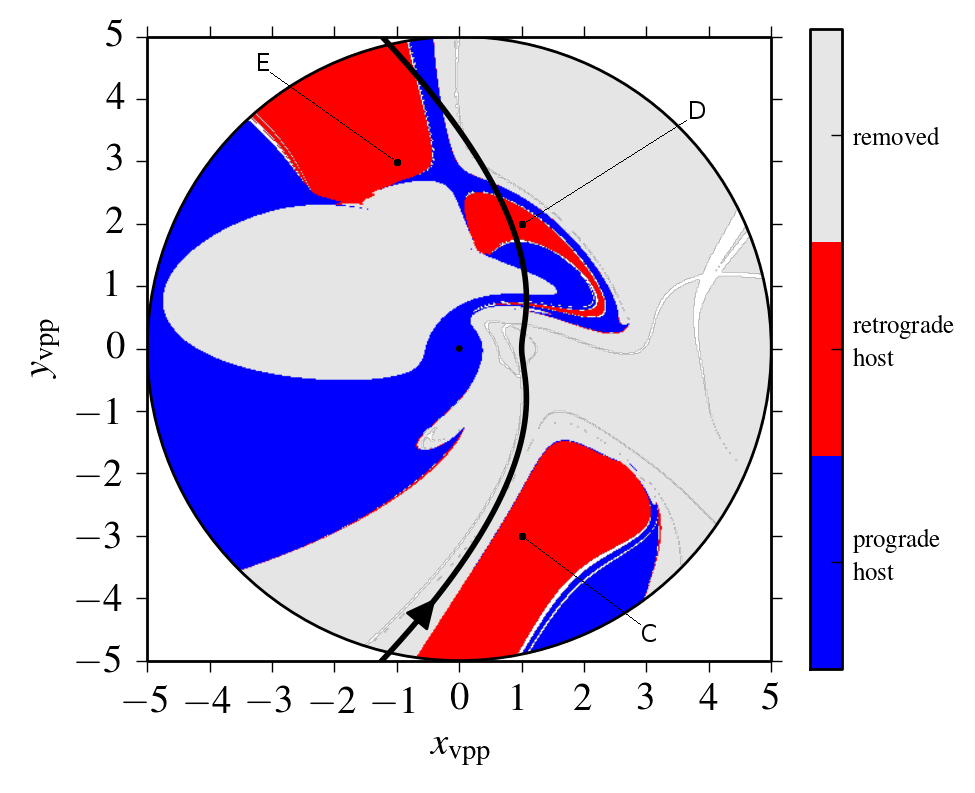}
      \end{minipage}
    \end{subfigure}
  \end{minipage}
  \begin{minipage}[t]{\hsize}
    \vspace{0pt}
    \begin{subfigure}[t]{\textwidth}
      \begin{minipage}[t]{0.04\textwidth}
        \vspace{0pt}
        \caption{}\label{fig:map_20}
      \end{minipage}
      \hfill
      \begin{minipage}[t]{0.9\textwidth}
        \vspace{0pt}
        \includegraphics[width=\textwidth]{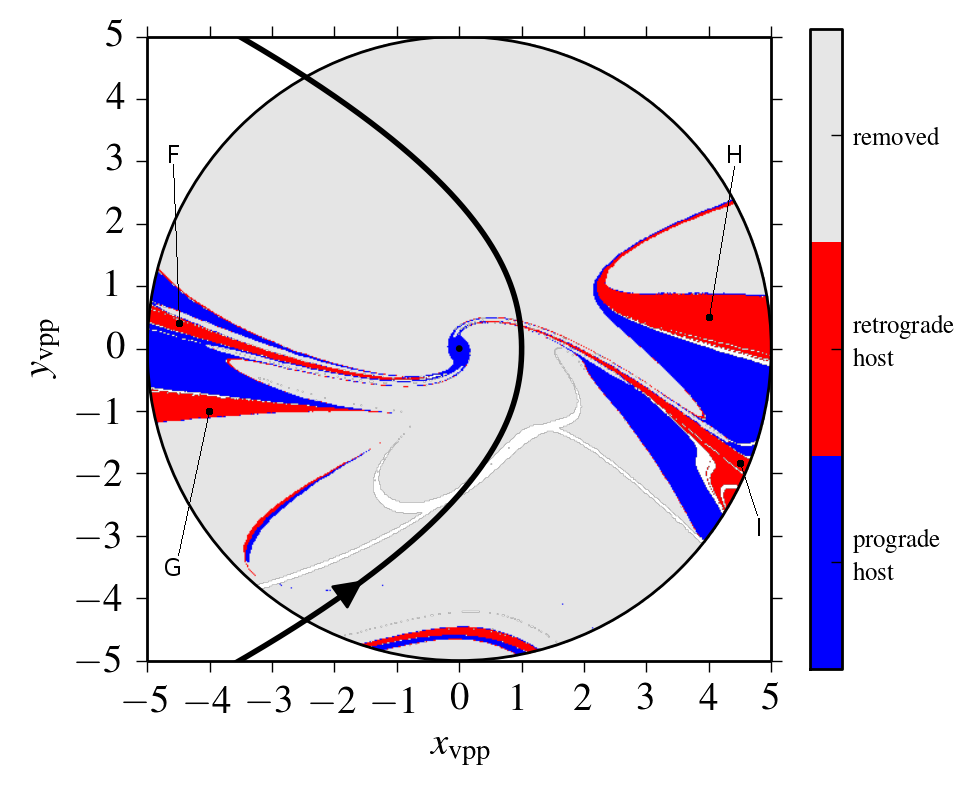}
      \end{minipage}
    \end{subfigure}
  \end{minipage}
  \caption{VPP maps showing the regions from which particles end up on retrograde orbits around the host after a prograde, coplanar fly-by for mass ratios of $m = 0.1$ ({\bf a}),
    $m = 1.0$ ({\bf b}), and \mbox{$m = 20.0$ ({\bf c})}.
    The thick black lines show the interaction orbit of the perturber as described in \citet{2017A&A...599A..91B}.
  }
  \label{fig:maps}
\end{figure}

The production of retrograde objects in retrograde, coplanar fly-bys (Table~\ref{tab:RC_probability_total}, $i = 180^{\circ}$) seems relatively straightforward.
Because the particles are attracted by the perturber and thus dragged along its orbit,
it was to be expected to find some particles finally on orbits with the same orientation of angular momentum as the perturber.
The probability for this case increases with the mass ratio.

\begin{figure}[ht!]
  \centering
  \begin{minipage}[t]{\hsize}
    \vspace{0pt}
    \begin{subfigure}[t]{\textwidth}
      \begin{minipage}[t]{0.04\textwidth}
        \vspace{0pt}
          \caption{}\label{fig:A}
      \end{minipage}
      \hfill
      \begin{minipage}[t]{0.95\textwidth}
        \vspace{0pt}
        \centering
        \includegraphics[width=0.8\textwidth]{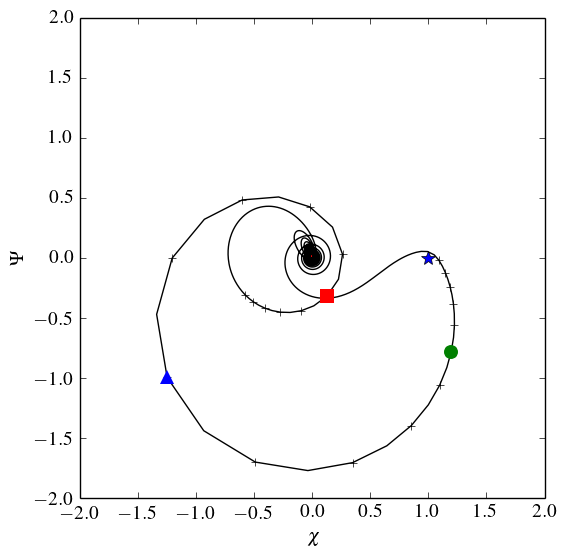}
        \end{minipage}
    \end{subfigure}
  \end{minipage}
  \begin{minipage}[t]{\hsize}
    \vspace{0pt}
    \begin{subfigure}[t]{\textwidth}
      \begin{minipage}[t]{0.04\textwidth}
        \vspace{0pt}
        \caption{}\label{fig:B}
      \end{minipage}
      \hfill
      \begin{minipage}[t]{0.95\textwidth}
        \vspace{0pt}
        \centering
        \includegraphics[width=0.8\textwidth]{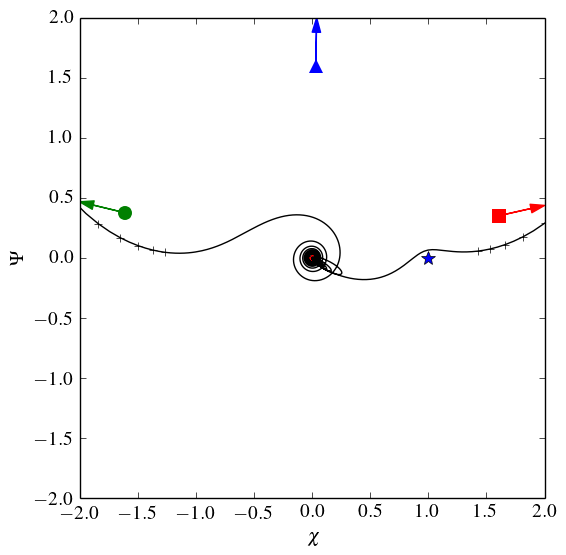}
      \end{minipage}
    \end{subfigure}
  \end{minipage}
  \caption{Interactions leading to retrograde orbits in the R3F for the cases\\
    \mbox{{\bf A (a)}: $i = 0$, $m = 0.1$, $x_{\mathrm{vpp}}  = 1$, $y_{\mathrm{vpp}}  = -3$}, and\\
    \mbox{{\bf B (b)}: $i = 0$, $m = 0.1$, $x_{\mathrm{vpp}}  = 0$, $y_{\mathrm{vpp}}  = 4.1$}.\\
    The green circles, blue triangles, and red squares mark the positions of the particles for $t = -t_{\mathrm{slr}}$, $t= 0$, and $t = t_{\mathrm{slr}}$, respectively.
    The definition of the Cartesian coordinates of the R3F, $\chi$ and $\Psi$,
    can be found in Sect.~\ref{sec:more_rotating} in the Appendix.
  }
  \label{fig:A_B}
\end{figure}

However, the production of retrograde objects in prograde coplanar fly-bys (Table~\ref{tab:RC_probability_total}, $i = 0^{\circ}$ ),
especially even with a significantly higher probability than in retrograde coplanar fly-bys in the equal-mass case, is rather surprising.
Therefore, we now more closely examine the particle trajectories that end in retrograde orbits around the host
in the case of prograde, coplanar perturbations.
Using VPP maps like in \citet{2017A&A...599A..91B}, we determine the initial conditions of these trajectories.

\subsection{Particle trajectories ending in retrograde orbits}
\label{sec:res_trajectories}

Figure~\ref{fig:maps} shows VPP maps for the region with \mbox{$r_{\mathrm{vpp}}  < 5$} perturbed by prograde, coplanar ($i = 0^{\circ}$) fly-bys with the three sample mass-ratios.
The colours in the maps indicate the final fate of the particles after the fly-by depending on their VPPs.
Particles with VPPs in the blue and red regions remain bound to the host.
Particles from the blue regions are finally on prograde orbits, and particles from the red regions on retrograde orbits.
Particles from the grey regions are removed from the host, either by being captured by the perturber or by becoming unbound.
The thick black lines show the interaction orbit of the perturber \citep[see][]{2017A&A...599A..91B}.

Figure~\ref{fig:maps} shows that whereas for fly-bys of a low-mass perturber most particles remain on prograde orbits, the situation is different for $m = 1.0$ and $m = 20.0$.
In the following, a representative sample trajectory is shown in the R3F
for each (major) region from which particles end up on retrograde orbits around the host.

\subsubsection{Low-mass perturbation}

For the mass-ratio of $m = 0.1$,
there are two different mechanisms that cause particles to be perturbed onto retrograde orbits around their host star.
The interactions for these cases are shown in Fig.~\ref{fig:A_B}.
The initial conditions for the interactions shown in Fig.~\ref{fig:A_B} are shown in Figure~\ref{fig:map_0.1} by the black dots labelled {\bf A} and {\bf B}.
All particles from the red regions around these dots undergo similar interactions with host and perturber.

\begin{figure*}[ht!]
  \begin{minipage}[t]{\hsize}
    \centering
    \begin{minipage}[t]{0.49\hsize}
      \vspace{0pt}
      \begin{subfigure}[t]{\textwidth}
        \begin{minipage}[t]{0.04\textwidth}
          \vspace{0pt}
          \caption{}\label{fig:F}
        \end{minipage}
        \hfill
        \begin{minipage}[t]{0.95\textwidth}
          \vspace{0pt}
          \includegraphics[width=0.8\textwidth]{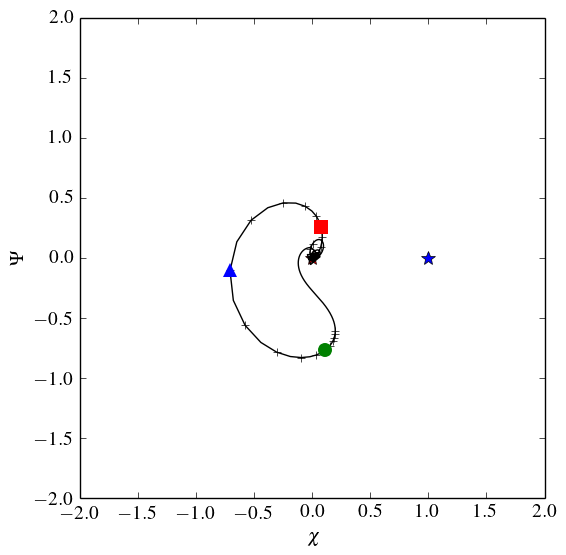}
        \end{minipage}
      \end{subfigure}
    \end{minipage}
    \begin{minipage}[t]{0.49\hsize}
      \vspace{0pt}
      \begin{subfigure}[t]{\textwidth}
        \begin{minipage}[t]{0.04\textwidth}
          \vspace{0pt}
          \caption{}\label{fig:G}
        \end{minipage}
        \hfill
        \begin{minipage}[t]{0.95\textwidth}
          \vspace{0pt}
          \includegraphics[width=0.8\textwidth]{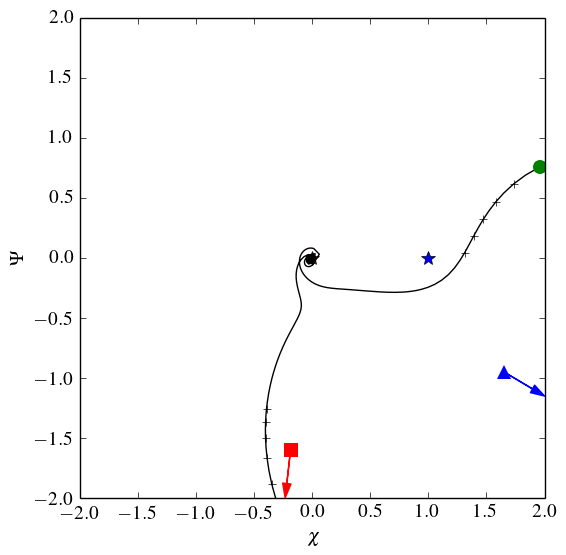}
        \end{minipage}
      \end{subfigure}
    \end{minipage}
  \end{minipage}
  \begin{minipage}[t]{\hsize}
    \centering
    \begin{minipage}[t]{0.49\hsize}
      \vspace{0pt}
      \begin{subfigure}[t]{\textwidth}
        \begin{minipage}[t]{0.04\textwidth}
          \vspace{0pt}
          \caption{}\label{fig:H}
        \end{minipage}
        \hfill
        \begin{minipage}[t]{0.95\textwidth}
          \vspace{0pt}
          \includegraphics[width=0.8\textwidth]{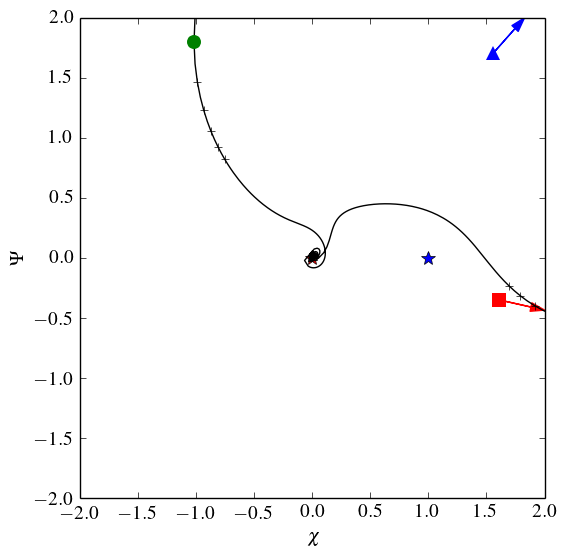}
        \end{minipage}
      \end{subfigure}
    \end{minipage}
    \begin{minipage}[t]{0.49\hsize}
      \vspace{0pt}
      \begin{subfigure}[t]{\textwidth}
        \begin{minipage}[t]{0.04\textwidth}
          \vspace{0pt}
          \caption{}\label{fig:I}
        \end{minipage}
        \hfill
        \begin{minipage}[t]{0.95\textwidth}
          \vspace{0pt}
          \includegraphics[width=0.8\textwidth]{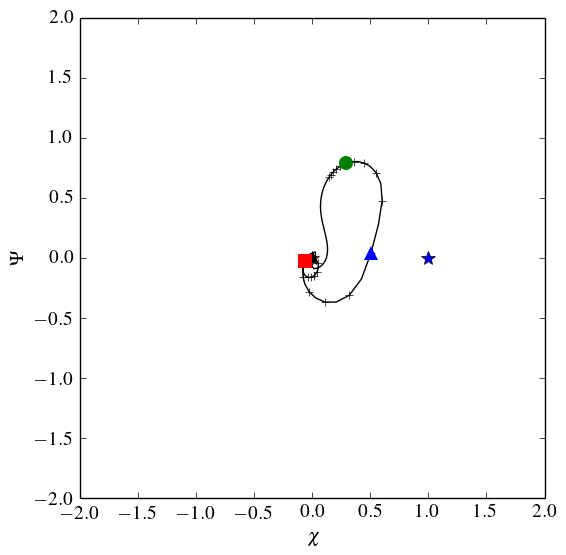}
        \end{minipage}
      \end{subfigure}
    \end{minipage}
  \end{minipage}
  \caption{Interactions leading to retrograde orbits in the R3F for the cases\\
    \mbox{{\bf F (a)}: $i = 0$, $m = 20.0$, $x_{\mathrm{vpp}}  = -4.5$, $y_{\mathrm{vpp}}  = 0.4$},
    \mbox{{\bf G (b)}: $i = 0$, $m = 20.0$, $x_{\mathrm{vpp}}  = -4$, $y_{\mathrm{vpp}}  = -1$},\\
    \mbox{{\bf H (c)}: $i = 0$, $m = 20.0$, $x_{\mathrm{vpp}}  = 4$, $y_{\mathrm{vpp}}  = 0.5$},
    \mbox{{\bf I (d)}: $i = 0$, $m = 20.0$, $x_{\mathrm{vpp}}  = 4.5$, $y_{\mathrm{vpp}}  =-1.8$}.\\
    The green circles, blue triangles, and red squares mark the positions of the particles for $t = -t_{\mathrm{slr}}$, $t= 0$, and $t = t_{\mathrm{slr}}$, respectively.
    The definition of the Cartesian coordinates of the R3F, $\chi$ and $\Psi$,
    can be found in Sect.~\ref{sec:more_rotating} in the Appendix.
  }
  \label{fig:F_G_H_I}
\end{figure*}

Figure~\ref{fig:A} shows the interaction for a particle with VPPs of $x_{\mathrm{vpp}}  = 1$, $y_{\mathrm{vpp}}  = -3$.
It can be seen that the particle moves counter clockwise around the host while the perturber approaches,
then it undergoes a close clockwise interaction with the perturber before $t = -t_{\mathrm{slr}}$ (particle position at $t = -t_{\mathrm{slr}}$ indicated by the green dot).
We note that the position of the perturber at $t = -t_{\mathrm{slr}}$ in normal space would be (0,\,-2).
When the perturber is in pericentre at $t= 0$ (particle position at that time indicated by blue triangle), the particle is on a retrograde orbit around the host,
passing the host on the opposite side from the perturber.
Shortly before $t = t_{\mathrm{slr}}$ (particle position at that time indicated by the red square), the particle passes between host and perturber and remains on a retrograde,
elliptical orbit around the host while the perturber departs (indicated by the elongated spiral movement of the particle towards the host).

The particles from region {\bf B} undergo a different series of interactions,
as shown in Fig.~\ref{fig:B} for a particle with VPPs of $x_{\mathrm{vpp}}  = 0$, $y_{\mathrm{vpp}}  = 4.1$.
Because of the strong zoom, the coloured reference points are outside the shown region. Their positions are approximately indicated by the arrows.
At $t = -t_{\mathrm{slr}}$ (indicated by the green dot and arrow), the particle is almost at the opposite side of the host from the perturber, with approximately twice the distance from the host as the perturber.
Because of the initial orbital radius of the particle of $r_{\mathrm{init}}  = 4.1$,
the particle pursues its orbit relatively unperturbed, while the perturber passes pericentre and moves to $SLR$.
After $t = 2\, t_{\mathrm{slr}}$, the perturber catches up with the particle and follows it roughly on its orbit.
During that time, the perturber attracts the particle, decelerates it, and accelerates it again into the opposite direction.
By that time, the influence of the perturber on the particle is no longer significant.
In contrast to the particles from region {\bf A}, the particles from region {\bf B}
become retrograde in an interaction with the perturber while the perturber is already departing from pericentre.
As shown above, the probability for a particle with \mbox{$0.1 < r_{\mathrm{init}}  < 10.0$} to undergo an interaction of type {\bf A} or {\bf B} with host and perturber is approximately $3\%$,
if all $r_{\mathrm{init}} $ between $0.1$ and $10.0$ are equally likely.

\subsubsection{Equal-mass perturbation}

For the case of an equal-mass perturber, the regions from 
which particles are excited onto retrograde orbits are shown in Fig.~\ref{fig:map_1}.
The interactions for this case are very similar to the interactions for case $m = 0.1$.
The interactions for region {\bf C} are similar to those for region {\bf A}.
The only significant difference is that the particle does not pass between host and perturber after the pericentre passage of the perturber, but passes the perturber outside its orbit.
The interactions for regions {\bf D} and {\bf E} are very similar to those for region {\bf B}.
The only significant differences here are the scales.
For the sake of completeness, the interactions in the R3F are shown together with the full descriptions in Sect.~\ref{sec:r3f_m1} in the Appendix.
The total probability for a particle to move on a trajectory similar to the cases {\bf C}, {\bf D}, or {\bf E} is approximately $13\%$, as shown above.

\subsubsection{High-mass perturbations}

In case of a high-mass perturber, the perturbation process is different.
Because of the high mass-ratio, the perturber rests at almost the centre of mass and the perturbation can be regarded as the host moving on a parabolic orbit around the perturber.
Therefore, the interactions of the particles with host and perturber are different from the low-mass and equal-mass case \citep[see also][]{2017A&A...599A..91B}.

Since the host moves around the almost resting perturber, the particles from the left half of Fig.~\ref{fig:map_20}, and especially the cases {\bf F} and {\bf G},
move approximately ahead of the host while approaching pericentre.
Figures~\ref{fig:F} and \ref{fig:G} show the interactions for these cases in the rotating reference frame.

For case {\bf F}, the particle moves slightly ahead of the host with a slightly larger distance from the perturber than the host.
When the host is at pericentre (particle position denoted by the blue triangle in Fig.~\ref{fig:F}), the particle has almost twice the distance from the perturber than the host.
Therefore, the host passes between particle and perturber.
Between $t = 0$ and $t = t_{\mathrm{slr}}$, the particle follows the host and catches up with it.
Shortly after $t = t_{\mathrm{slr}}$, it passes between host and perturber and then enters an eccentric, retrograde orbit around the host.

The particle for case {\bf G} moves ahead of the host while approaching pericentre and passes the perturber already before $t = -2 t_{\mathrm{slr}}$.
When the host passes $SLR$ (particle position indicated by the red square), the particle is far ahead of the host.
Then the host catches up with the particle, however, and passes between particle and perturber, forcing the particle into a retrograde orbit.

The particles from the right half of Fig.~\ref{fig:map_20}, and especially cases {\bf H} and {\bf I}, move approximately behind the host while approaching pericentre.
Figures~\ref{fig:H} and \ref{fig:I} show the interactions for these cases in the rotating reference frame.

In case {\bf H}, the particle moves far behind the host while approaching pericentre.
When the host passed $SLR$ (particle position indicated by red square), the particle did not yet pass the line between perturber and $-SLR$.
After moving around the perturber, the particle follows the host on a similar trajectory and with a similar velocity, but heading slightly left to the host.
After a relatively long time, the particle settles into a retrograde orbit around the host.

For case {\bf I}, the particle moves closely behind the host while approaching pericentre.
When the host is at pericentre, the particle is located approximately between host and perturber, thus moving on a closer and faster orbit around the perturber.
The particle moves then ahead of the host for a short time, increases its distance to the perturber and thereby crosses the hosts orbit.
When the host reaches $SLR$, the host passes between particle and perturber, very close to the particle, forcing the particle into a tight, retrograde orbit around itself.
The probability for a particle to move on a trajectory similar to cases {\bf F} to {\bf I} is approximately $8\%$.

\begin{figure}[ht!]
  \centering
  \begin{minipage}[t]{\hsize}
    \vspace{0pt}
    \begin{subfigure}[t]{\textwidth}
      \begin{minipage}[t]{0.04\textwidth}
        \vspace{0pt}
          \caption{}\label{fig:CO_in_coplanar_prograde_0.1}
      \end{minipage}
      \hfill
      \begin{minipage}[t]{0.95\textwidth}
        \vspace{0pt}
        \includegraphics[width=\textwidth]{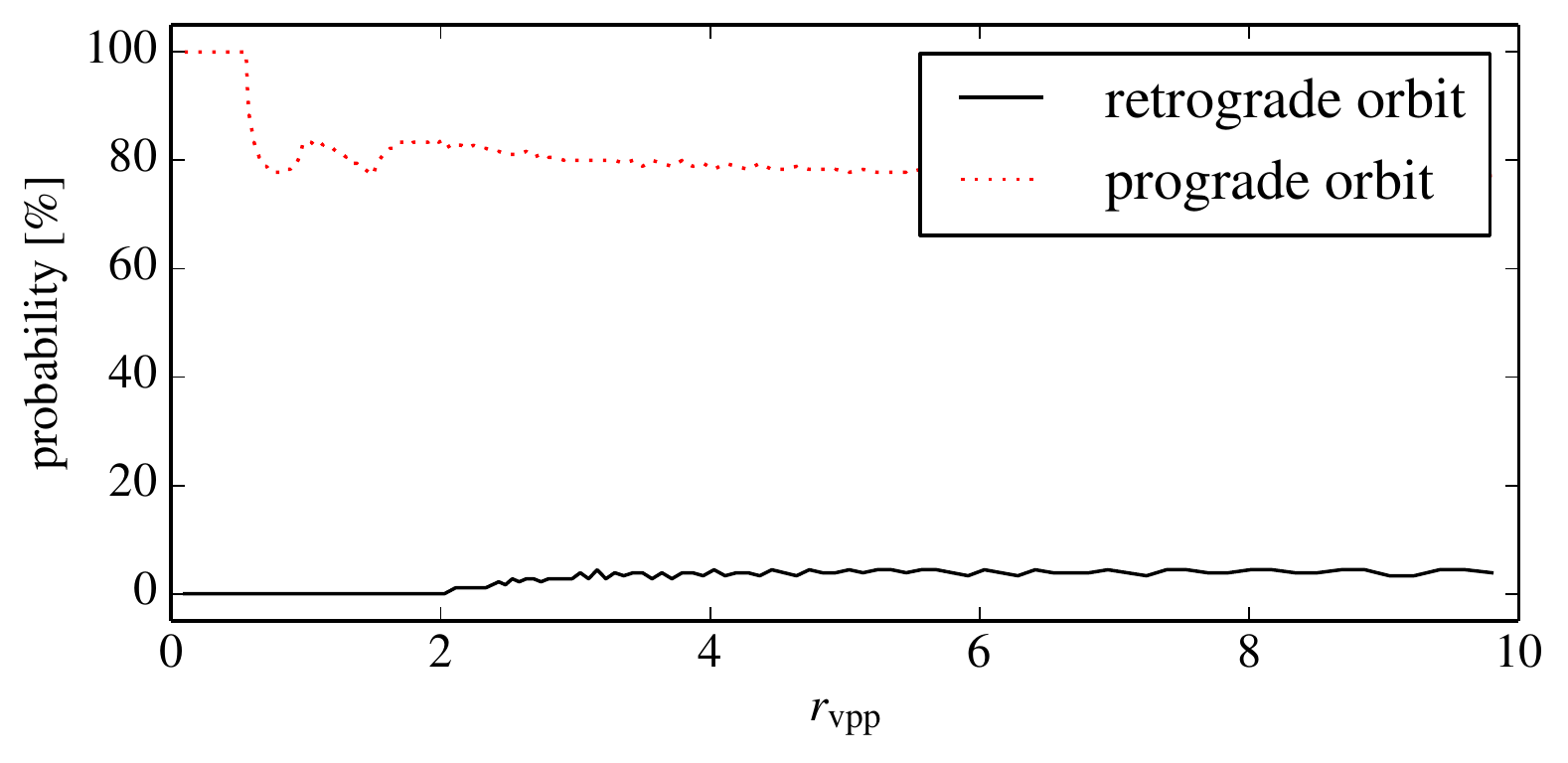}
        \end{minipage}
    \end{subfigure}
  \end{minipage}
  \begin{minipage}[t]{\hsize}
    \vspace{0pt}
    \begin{subfigure}[t]{\textwidth}
      \begin{minipage}[t]{0.04\textwidth}
        \vspace{0pt}
        \caption{}\label{fig:CO_in_coplanar_prograde_1}
      \end{minipage}
      \hfill
      \begin{minipage}[t]{0.95\textwidth}
        \vspace{0pt}
        \includegraphics[width=\textwidth]{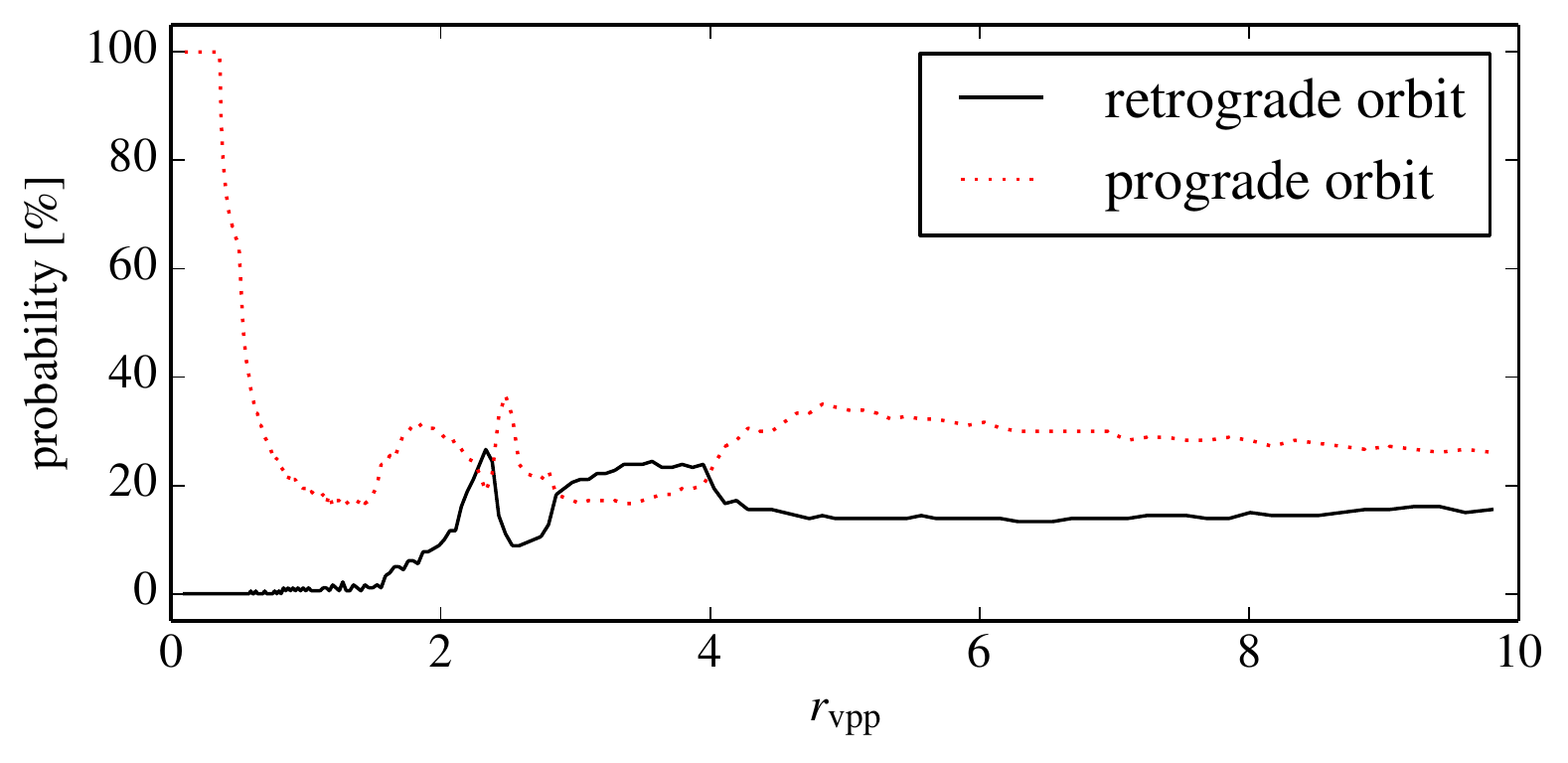}
      \end{minipage}
    \end{subfigure}
  \end{minipage}
  \begin{minipage}[t]{\hsize}
    \vspace{0pt}
    \begin{subfigure}[t]{\textwidth}
      \begin{minipage}[t]{0.04\textwidth}
        \vspace{0pt}
        \caption{}\label{fig:CO_in_coplanar_prograde_20}
      \end{minipage}
      \hfill
      \begin{minipage}[t]{0.95\textwidth}
        \vspace{0pt}
        \includegraphics[width=\textwidth]{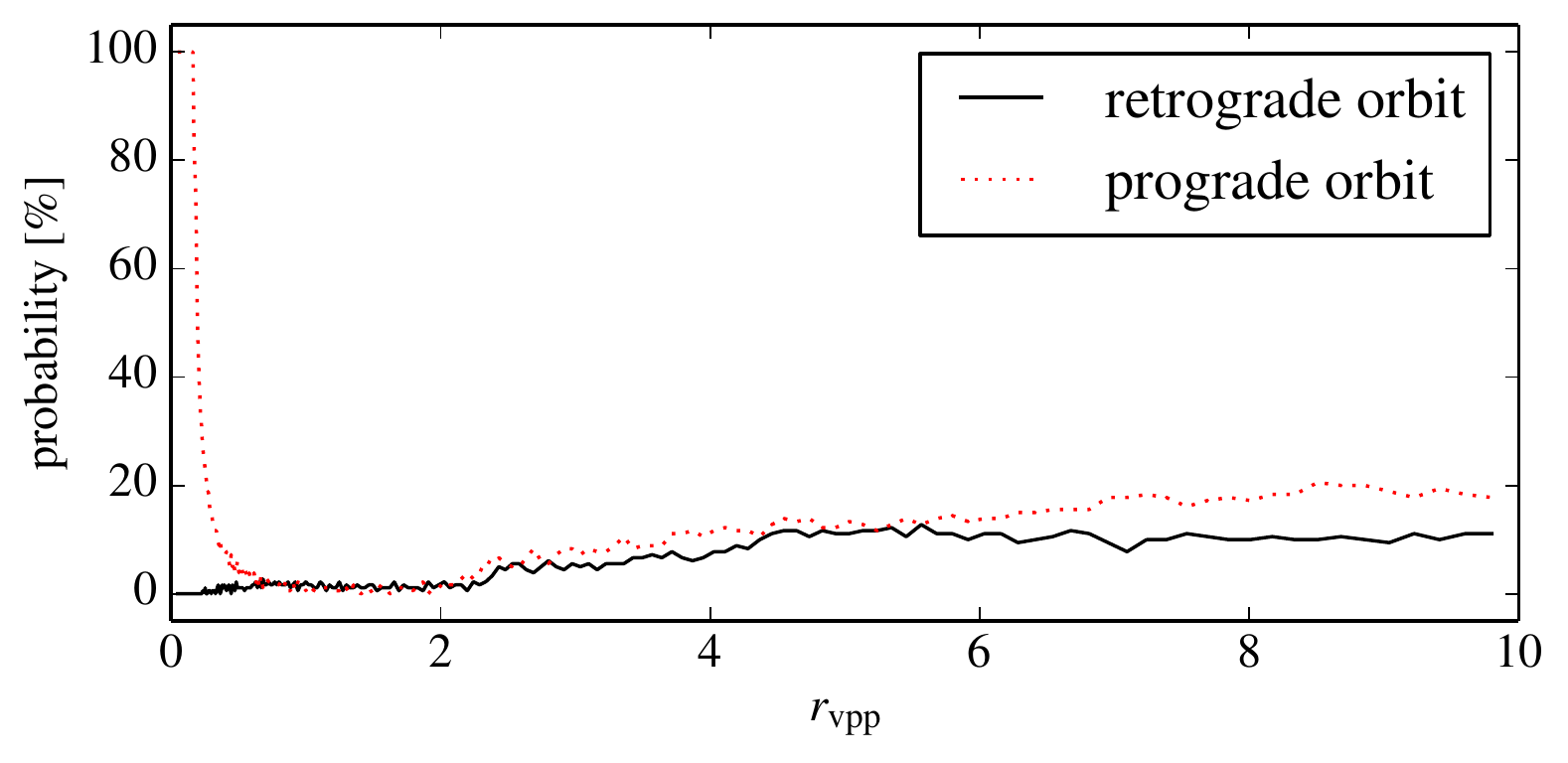}
      \end{minipage}
    \end{subfigure}
  \end{minipage}
  \caption{Probability for a planet (or disc material) to end up on a prograde or retrograde orbit
    in a coplanar, prograde encounter with a perturber with mass-ratio $m = 0.1$ ({\bf a}), $m = 1.0$ ({\bf b}), and $m = 20.0$ ({\bf c}) as a function of $r_{\mathrm{vpp}} $.
  }
  \label{fig:CO_in_coplanar_prograde}
\end{figure}

\subsection{Radial probabilities for retrograde orbits}
\label{sec:res_probability}

The fate of a particle depends on both the initial orbital radius relative to the pericentre distance of the perturber orbit, $r_{\mathrm{init}} $,
and its initial angle $\theta_{\mathrm{init}} $. Here, we restrict our investigation to
material that becomes counter-orbiting or retrograde as a function of $r_{\mathrm{init}} $,
starting with an examination of the case of a prograde, coplanar perturbation.

Figure~\ref{fig:CO_in_coplanar_prograde} shows the probabilities for an object initially on a circular orbit around the host to end up in a prograde or retrograde orbit around the host
depending on its initial orbital radius around the host, when perturbed by a coplanar, prograde fly-by.
Like before, for a mass ratio of $m = 0.1$ (see Fig.~\ref{fig:CO_in_coplanar_prograde_0.1}), the probability for an object to end up in a retrograde orbit around the host is in general relatively small.
For \mbox{$r_{\mathrm{vpp}}  \lesssim 2$}, this probability is zero, for \mbox{$2 \lesssim r_{\mathrm{vpp}}  \lesssim 10$}, it is below $5\%$.
In contrast, the probability to remain in a prograde orbit around the host is $100\%$ for $r_{\mathrm{vpp}}  \lesssim 0.6$ and $\approx 80 \pm 5 \%$ for $r_{\mathrm{vpp}}  \gtrsim 0.6$.

For a mass ratio of $m = 1$ (see Fig.~\ref{fig:CO_in_coplanar_prograde_1}), the probability to end up in a retrograde orbit around the host is much higher than for $m = 0.1$
and increases for large $r_{\mathrm{vpp}} $.
It is zero for $r_{\mathrm{vpp}}  \lesssim 0.6$ and $< 3\%$ for $0.6 \lesssim r_{\mathrm{vpp}}  \lesssim 1.6$.
For $r_{\mathrm{vpp}}  \gtrsim 4$, it is $15 \pm 2\%$, and between $r_{\mathrm{vpp}}  \gtrsim1.6$ and $r_{\mathrm{vpp}}  \lesssim 4$, it is up to $25\%$.
The probability for a prograde orbit, in contrast, is between $25$ and $35 \%$ for $r_{\mathrm{vpp}}  \gtrsim 4$.
For $r_{\mathrm{vpp}} $ between $3$ and $4$, the probability for a prograde orbit is even below the probability for a retrograde orbit.
We note that the sum does not reach 100\% because matter can also become unbound and the fraction of unbound matter becomes larger for higher mass ratios.

For a mass ratio of $m = 20$ (see Fig.~\ref{fig:CO_in_coplanar_prograde_20}), the probability to end up in a retrograde orbit around the host is again different.
For $r_{\mathrm{vpp}}  \lesssim 2$, the probability is $< 2\%$.
For $2 \lesssim  r_{\mathrm{vpp}}  \lesssim 4.5$, the probability increases to $\approx 10\%$, where it remains up to $r_{\mathrm{vpp}}  = 10$.
In contrast, the probability to end up in a prograde orbit around the host is $100\%$ for $r_{\mathrm{vpp}}  \lesssim 0.2$.
Then it falls to $\approx 2 \pm 1\%$ for $r_{\mathrm{vpp}}  \approx 0.6$ and remains there until $r_{\mathrm{vpp}}  \approx 2$.
For $r_{\mathrm{vpp}}  \gtrsim 2$ the probability rises almost linearly to $\approx 20 \pm 2\%$ for $r_{\mathrm{vpp}}  = 10$.

\begin{figure}[t!]
  \centering
  \begin{minipage}[t]{\hsize}
    \vspace{0pt}
    \begin{subfigure}[t]{\textwidth}
      \begin{minipage}[t]{0.04\textwidth}
        \vspace{0pt}
          \caption{}\label{fig:inclination_averaged_CO_probs}
      \end{minipage}
      \hfill
      \begin{minipage}[t]{0.95\textwidth}
        \vspace{0pt}
        \includegraphics[width=\textwidth]{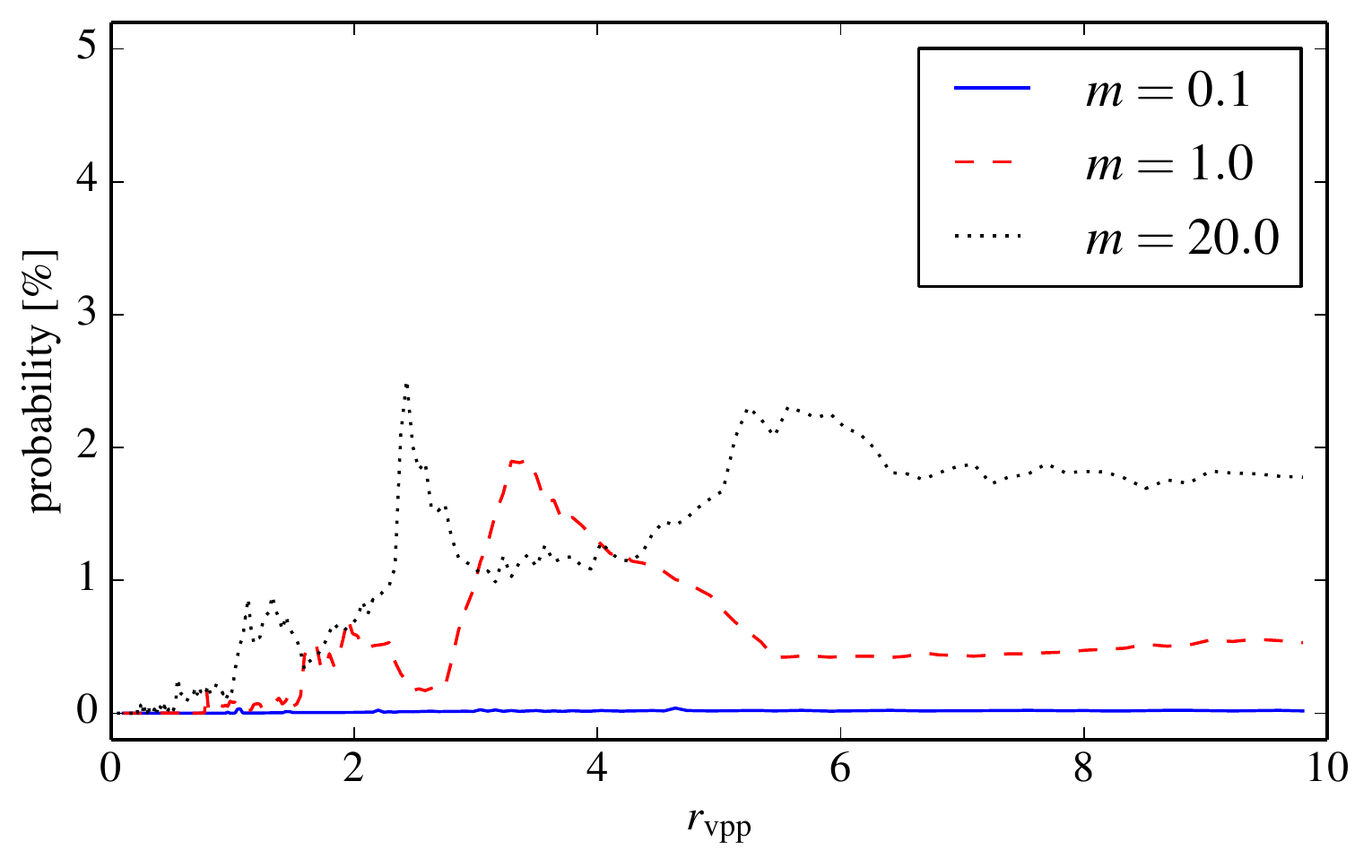}
        \end{minipage}
    \end{subfigure}
  \end{minipage}
  \begin{minipage}[t]{\hsize}
    \vspace{0pt}
    \begin{subfigure}[t]{\textwidth}
      \begin{minipage}[t]{0.04\textwidth}
        \vspace{0pt}
        \caption{}\label{fig:inclination_averaged_RETROGRADE_probs}
      \end{minipage}
      \hfill
      \begin{minipage}[t]{0.95\textwidth}
        \vspace{0pt}
        \includegraphics[width=\textwidth]{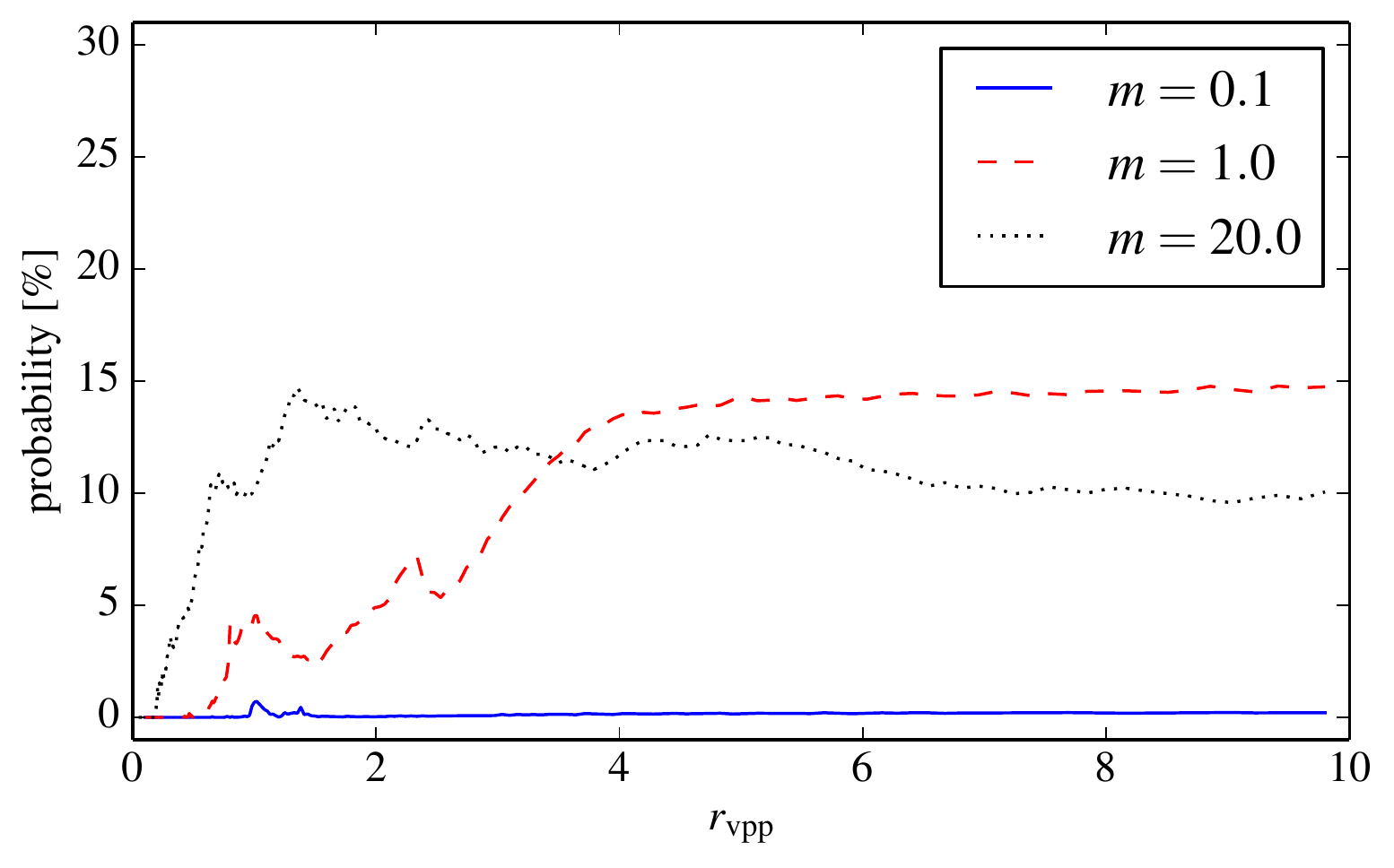}
      \end{minipage}
    \end{subfigure}
  \end{minipage}
  \caption{Probability for a planet (or disc material) to become
    {\bf a)} counter-orbiting ($\varphi > 170 ^{\circ}$) and {\bf b)} retrograde ($\varphi > 90 ^{\circ}$),
    averaged over all parabolic perturbations with the inclinations we used in Sect.~\ref{sec:total_probabilities} for the three mass-ratios as a function of $r_{\mathrm{vpp}} $.
  }
  \label{fig:inclination_averaged_probabilities}
\end{figure}

For the mass-ratios illustrated, only the low-mass case has a significantly lower chance of producing retrogradely orbiting particles than prograde particles
over the radial range illustrated.
For the equal-mass case and the high-mass case, the probabilities are of the same order of magnitude outside $r_{\mathrm{vpp}}  \gtrsim 0.5 \text{--} 1.0$.
For the equal-mass case, there is even a radial range for which the probability to form retrogradely orbiting particles is higher than to form progradely orbiting particles ($3 \lesssim r_{\mathrm{vpp}}  \lesssim 4$).
For an object with a certain initial orbital radius $r$, this means that the probability to end up on a retrograde orbit around the host is only significantly non-zero when the
pericentre distance of the perturber orbit is $r_{\mathrm{p,peri}}  \lesssim r/2$.
In general, the probability increases for smaller pericentre distances, with a maximum for $m = 1.0$ being reached at a pericentre distance $r_{\mathrm{p,peri}}  \approx r/3$.

However, the probabilities shown in Fig.~\ref{fig:CO_in_coplanar_prograde} are only valid for perturbations by a coplanar, prograde perturber --
hence 'retrograde' and 'counter-orbiting' are equivalent.
In Fig.~\ref{fig:RC_vs_inclination} we have already shown that the probability to produce counter-orbiting objects by a perturbation is significantly
non-zero only for inclinations of $i= 0 ^{\circ}$ and $i = 180 ^{\circ}$.
Therefore, we now examine the inclination-averaged probability to form counter-orbiting and retrograde particles.

The inclination-averaged probabilities to produce counter-orbiting objects by a fly-by are shown in Fig.~\ref{fig:inclination_averaged_CO_probs} for the three mass-ratios.
For this plot, data like those shown in Fig.~\ref{fig:CO_in_coplanar_prograde} have been integrated over the inclinations of the perturber orbit
we used in Sect.~\ref{sec:total_probabilities}
using the trapezoidal rule and were normalised afterwards.
We recall that we only consider arguments of periapsis of $\omega = 0$ throughout this paper.

\begin{table*}[t]
  \caption{Cluster models for estimating fly-by probabilities.}
  \label{tab:cluster_properties}
  \centering
  \begin{tabular}{lllllll}
    \toprule
    Model name  & Resembling & $m_{\mathrm{cl}}$ [$M_{\odot}$] & $r_{\mathrm{h}}$ [pc] & $\bar{m}_*$ [$M_{\odot}$] & $N_{\mathrm{solar}}$ & Ref\\
    \midrule
    C1          & ONC        &          \hphantom{1}2000 &        0.8 &           0.7 & \hphantom{00}60 & [1] \\
    C2          & Arches     &                     50000 &        0.2 &           1.6 &            3300 & [2] \\
    \bottomrule
  \end{tabular}
  \tablefoot{Here, $m_{\mathrm{cl}}$ is the total cluster mass and $r_{\mathrm{h}}$ is the half-mass radius.
    The mean stellar mass, $\bar{m}_*$, and the approximate number of solar mass stars, $N_{\mathrm{solar}}$, were calculated
    based on the mass functions for the respective cluster.
    Model C2 resembles the ``initial'' Arches, not the contemporary Arches.
  }
  \tablebib{[1]~\citet{1998ApJ...492..540H};
    [2]~\citet{2010MNRAS.409..628H}}
\end{table*}

Figure~\ref{fig:inclination_averaged_CO_probs} shows that the averaged probability to produce a counter-orbiting object in a perturbation by a perturber
with a mass-ratio of $m = 0.1$ is negligible for the whole considered radial range.
For $m = 1$, this probability is $\approx 0\%$ for $r_{\mathrm{vpp}}  \lesssim 1.5$, has a maximum of $\approx 2\%$ for $r_{\mathrm{vpp}}  \approx 3.5$,
and is almost constant $\approx 0.5\%$ for $r_{\mathrm{vpp}}  \gtrsim 5.5$.
For $m = 20$, the probability is $\approx 0\%$ for $r_{\mathrm{vpp}}  \lesssim 1$, increases almost linearly up to $\approx 2\%$ at $r_{\mathrm{vpp}}  \approx 5$
with a small peak of almost $3\%$ for $r_{\mathrm{vpp}}  \approx 2.5$, and is almost constant $\approx 2\%$ for $r_{\mathrm{vpp}}  \gtrsim 5$.

Figure~\ref{fig:inclination_averaged_RETROGRADE_probs} in contrast shows the probabilities of matter to become retrograde in general ($\varphi > 90 ^{\circ}$).
For $m = 0.1$, the probability to produce retrograde objects is also negligible over the full radial range.
Only at $r_{\mathrm{vpp}}  \approx 1$ is there a small maximum of $\approx 0.7\%$.
For $m = 1.0$, the probability is $\approx 0\%$ for $r_{\mathrm{vpp}}  \lesssim 0.5$ and then increases to $\approx 14\%$ at $r_{\mathrm{vpp}}  \approx 4$, where it remains until $r_{\mathrm{vpp}}  = 10$.
For $m = 20.0$, the probability is $\approx 0\%$ for $r_{\mathrm{vpp}}  \lesssim 0.2$, then increases to almost $15\%$ at $r_{\mathrm{vpp}}  \approx 1.3$, and finally decreases to $\approx 10\%$ at $r_{\mathrm{vpp}}  = 10$.

We note that in contrast to the probabilities in Sect.~\ref{sec:total_probabilities}, these probability distributions are independent of the orbital radius distribution.
The total probabilities from Sect.~\ref{sec:total_probabilities} can be obtained by convolving these probabilities with the orbital radius distribution.
We also note that the probabilities shown here are the probabilities for matter with a certain $r_{\mathrm{vpp}} $ to become counter-orbiting or retrograde during a certain fly-by.
They hold no information about the probability to find matter with a certain orbital radius after the perturbation.

In summary, fly-bys where the perturber is of similar or higher mass than the host star are generally those most likely to produce planets on retrograde orbits.
In these cases, there is even a non-negligible chance to produce counter-orbiting planets with this mechanism.
Generally, retrograde and counter-orbiting planets are more likely for high $r_{\mathrm{vpp}} $ values,
that is, for close fly-bys with a pericentre smaller than the perturbed particles orbits.

\section{Discussion}
\label{sec:discussion}

Here we studied whether stellar fly-bys are able to excite matter that initially orbits a star on prograde orbits to retrograde ones.
The numerical method we used is the same as in \cite{2017A&A...599A..91B},
therefore, the points of discussion for this method also apply here.
To summarise, they are the numerical integrator, the error tolerance of the integrator, and the start and end times of the simulations.
In contrast to most similar studies, we integrate the trajectory of each particle individually and in particular end the integration only when the particle has settled again into a stable orbit.
Therefore, the effect of the limitations listed above on the integrated trajectories can in general be considered comparable to or even smaller than in similar studies.
For further details, we refer to the discussion of \cite{2017A&A...599A..91B}.

In this study we consider only the gravitational influence of the two stars on the particles and integrate the trajectories individually.
We completely neglect any type of particle-particle interactions and especially viscous effects.
Therefore, the method and results can in principle be applied to three cases: (a) a debris disc, (b) the parameter space of possible planetary orbits, and (c) a protoplanetary disc.
In the case of the perturbation of a planetary system or a debris disc, neglecting viscous effects is certainly justified.
In the case of the perturbation of a protoplanetary disc, this simplification may only be justified for distant or grazing encounters, but not for penetrating ones.
Therefore, we wish to give an estimate of how common encounters are for which neglecting viscosity might be a problem.

Most stars are born in stellar clusters \citep[e.g.][]{2003ARA&A..41...57L},
and most of these stars are initially surrounded by protoplanetary discs.
Depending on the stellar density in the cluster, strong gravitational interactions between these young stars may be common.
It has been shown that such interactions may result in inclined discs if the perturber orbit is inclined relative to the disc \mbox{\citep[e.g.][]{1993MNRAS.261..190C,2016MNRAS.455.3086X}}.
Here we have demonstrated that even coplanar interactions might lead to the production of retrogradely orbiting protoplanetary disc matter.

As shown in Sect.~\ref{sec:results}, the production of counter-orbiting matter ($\varphi \approx 180^{\circ}$) is only likely
in prograde, coplanar encounters with $r_{\mathrm{p,peri}}  \lesssim 0.5\,r_{disc}$.
For typical sizes of protoplanetary discs in this evolutionary stage of a few $100$~au \citep[e.g.][]{2000AJ....119.2919B,2009ApJ...700.1502A},
this condition would be fulfilled by encounters with $r_{\mathrm{p,peri}}  \approx 50 \text{--} 100$~au.
It depends on the type of stellar group whether such encounters occur frequently (see estimate below).

In case of a high-mass perturber, only very little material might remain after such an encounter ($r_{\mathrm{p,peri}}  \lesssim 0.5\,r_{disc}$).
For lower mass-ratios ($m \lesssim 1$), the remaining prograde matter of the disc will likely dominate (compare Fig.~\ref{fig:CO_in_coplanar_prograde}).
For this case, our results are not directly applicable because of the neglected viscous effects.
When viscosity is taken into consideration, interactions between the prograde and retrograde components of the disc will dampen the retrograde component and leave only the prograde component,
now with less angular momentum than initially \citep[as has been shown e.g. by][]{2007A&A...462..193P}.

Clusters expand significantly within the first $10$~Myr of their development, therefore close interactions become rarer with higher cluster age.
Nevertheless, especially for stars in long-lived open clusters, there is a certain chance that they might still
have some close interactions.
At this stage, gravitational interactions with the other cluster members may influence the orbits of already formed planets.
Depending on the encounter parameters, these interactions might result in retrograde or even counter-orbiting planets.
The planetary orbits resulting from these interactions may afterwards be altered by long-term processes within the planetary system
\citep[e.g.][]{2005Icar..173..559M,2011MNRAS.411..859M,2014MNRAS.444.2808P}.

\begin{table}[t!]
  \caption{Approximate number of encounters of solar-mass stars per Myr depending on cluster model and periastron distance.}
  \label{tab:flyby_probabilities}
  \centering
  \begin{tabular}{lll}
    \toprule
    $r_{\mathrm{p,peri}} $ [au]  & C1    &               C2 \\
    \midrule
               1000 & 6.7   &            26000 \\
    \hphantom{0}100 & 0.7   & \hphantom{0}2600 \\
    \hphantom{00}10 & 0.07  & \hphantom{00}260 \\
    \hphantom{000}1 & 0.007 & \hphantom{000}26 \\
    \bottomrule
  \end{tabular}
\end{table}

The overall probability to produce retrograde or counter-orbiting matter by stellar interactions in stellar clusters in both above-mentioned stages of cluster evolution is very difficult to estimate.
For a rough estimate, Table~\ref{tab:cluster_properties} lists sample properties for an ONC-like and an Arches-like cluster model.
These models represent two differently dense cluster environments.
Based on these properties, the approximate number of encounters between two solar-mass stars per Myr can be estimated depending on the periastron distance using Eq.~(3) of \citet{2007MNRAS.378.1207M}
(see Table~\ref{tab:flyby_probabilities}).
To obtain these values, the number of encounters per solar-mass star per Myr has been multiplied with the approximate number of stars with $0.9\,M_{\odot}  < m < 1.1\,M_{\odot} $ in these clusters,
$N_{\mathrm{solar}}$, from Table~\ref{tab:cluster_properties}.
Depending on the size of a protoplanetary disc or an already formed planetary system, fly-bys with $100$, $10$, or $1$~au would fulfil
the criterion that $r_{\mathrm{p,peri}}  \lesssim 0.5\,r_{disc}$ or the orbital radius of a planet.
When this criterion is fulfilled, the excitation of retrograde and / or counter-orbiting matter is possible in general.
The table shows that only for the Arches model, such fly-bys occur reasonably frequently.
In an Arches-like cluster, about $\approx 250$ equal-mass fly-bys with $r_{\mathrm{p,peri}}  \approx 10$~au per Myr can be expected.
For any cluster like the ONC or with an even lower density, such fly-bys are very unlikely.

If all of these fly-bys were parabolic, the
probability for the production of retrograde orbits ($\varphi > 90^{\circ}$) by such an encounter would be about $10\%$.
The probability for the production of counter-orbiting matter ($\varphi \approx 180^{\circ}$) by such fly-bys would be only about $1\%$.
In case of a high-mass perturber, the probabilities are slightly higher, in case of low-mass perturbers ($m \ll 1$), both probabilities are negligible.
When considering other than equal-mass encounters, the numbers of encounters per Myr are different.
Because of the higher number of low-mass stars, encounters between two of them are always more likely than between two solar-mass stars.

For a more detailed statement about the probability to produce retrogradely orbiting matter by stellar encounters in star clusters,
especially under consideration of fly-by eccentricities, mass ratios, possibly present residual gas, and the expansion of the cluster,
the analysis of the encounter history of the stars in a cluster as performed for instance by \citet{2006ApJ...642.1140O}, \citet{2014A&A...565A..32S}
, \citet{2015A&A...577A.115V}, and \citet{2016MNRAS.457..313P} has to be adopted.
Here, the inclinations between perturber orbit and disc have also to be considered \citep[e.g.][]{2001Icar..153..416K,2016A&A...594A..53B}.

\section{Practical consequences}
\label{sec:application}

\subsection{Application to the solar system}

\begin{table}[t!]
  \centering
  \begin{tabular}{rrrrl}
    \toprule
    $i$ [$^{\circ}$] & $e$ & $a$ [au] & \multicolumn{2}{c}{Designation / Name} \\
    \midrule
    140.8 & 0.97 & 309.0 & (336756) & 2010 NV$_{\mathrm{1}}$\\ 
    143.9 & 0.92 &  27.0 & & 2010 OR$_{\mathrm{1}}$\\
    143.9 & 0.99 & 432.6 & & 2010 BK$_{\mathrm{118}}$ \\ 
    144.0 & 0.86 &  73.8 & & 2016 NM$_{\mathrm{56}}$ \\ 
    146.3 & 0.91 &  32.0 & & 2010 CG$_{\mathrm{55}}$ \\
    146.9 & 0.96 &  62.1 & & 2012 HD$_{\mathrm{2}}$\\
    147.8 & 0.92 &  20.2 & & 2009 YS$_{\mathrm{6}}$\\
    148.4 & 0.85 &  11.2 & & 2016 VY$_{\mathrm{17}}$ \\
    150.2 & 0.94 &  42.8 & & 2006 EX$_{\mathrm{52}}$ \\
    151.8 & 0.47 &   8.1 & & 1999 LE$_{\mathrm{31}}$ \\
    152.3 & 0.66 &  13.0 & & 2016 JK$_{\mathrm{24}}$ \\
    152.4 & 0.98 & 185.5 & & 2017 CW$_{\mathrm{32}}$ \\
    154.8 & 0.97 &  79.5 & & 2013 LD$_{\mathrm{16}}$ \\
    156.5 & 0.94 &  23.9 & & 2010 EB$_{\mathrm{46}}$ \\
    157.5 & 0.90 &  37.1 & & 2015 XR$_{\mathrm{384}}$ \\
    158.5 & 0.90 &  23.6 & & 2000 HE$_{\mathrm{46}}$ \\
    159.1 & 0.85 &  14.6 & & 2015 XX$_{\mathrm{351}}$ \\
    159.2 & 0.47 &   5.5 & & 2017 BD$_{\mathrm{86}}$ \\
    160.0 & 0.88 &  29.6 & & 2012 TL$_{\mathrm{139}}$ \\
    160.4 & 0.90 &  23.9 & (20461) & Dioretsa \\
    163.0 & 0.38 &   5.1 & & 2015 BZ$_{\mathrm{509}}$ \\
    164.6 & 0.76 &   9.7 & & 2006 RJ$_{\mathrm{2}}$\\
    165.3 & 0.80 &   9.6 & & 2006 BZ$_{\mathrm{8}}$\\
    165.5 & 0.98 &  97.3 & & 2004 NN$_{\mathrm{8}}$\\
    165.6 & 0.40 &  10.9 & (459870) & 2014 AT$_{\mathrm{28}}$ \\ 
    170.3 & 0.57 &   8.1 & (330759) & 2008 SO$_{\mathrm{218}}$ \\
    170.8 & 0.87 &  32.4 & & 2014 CW$_{\mathrm{14}}$ \\
    171.0 & 0.96 &  67.1 & & 2016 EJ$_{\mathrm{203}}$ \\
    172.9 & 0.25 &   6.7 & (434620) & 2005 VD \\
    175.1 & 0.47 &   5.7 & & 2013 LA$_{\mathrm{2}}$\\
    \bottomrule
  \end{tabular}
  \caption{Solar system minor bodies with semi-major axes $a \gtrsim 5$~au and inclinations $i > 140^{\circ}$.
    Values taken from the IAU Minor Planet Center, http://www.minorplanetcenter.net.
  }
  \label{tab:minor_bodies}
\end{table}

\begin{figure}[t!]
  \centering
  \begin{minipage}[t]{\hsize}
    \vspace{0pt}
    \begin{subfigure}[t]{\textwidth}
      \begin{minipage}[t]{0.04\textwidth}
        \vspace{0pt}
          \caption{}\label{fig:e-a_0.1}
      \end{minipage}
      \hfill
      \begin{minipage}[t]{0.95\textwidth}
        \vspace{0pt}
        \includegraphics[width=\textwidth]{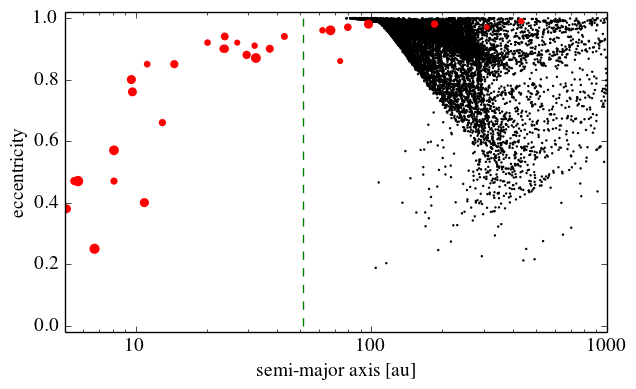}
        \end{minipage}
    \end{subfigure}
  \end{minipage}
  \begin{minipage}[t]{\hsize}
    \vspace{0pt}
    \begin{subfigure}[t]{\textwidth}
      \begin{minipage}[t]{0.04\textwidth}
        \vspace{0pt}
        \caption{}\label{fig:e-a_1.0}
      \end{minipage}
      \hfill
      \begin{minipage}[t]{0.95\textwidth}
        \vspace{0pt}
        \includegraphics[width=\textwidth]{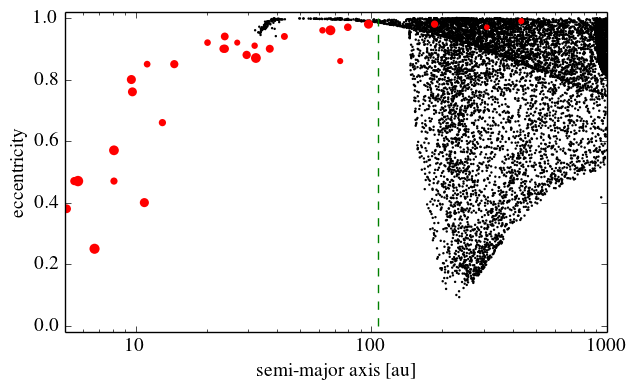}
      \end{minipage}
    \end{subfigure}
  \end{minipage}
  \begin{minipage}[t]{\hsize}
    \vspace{0pt}
    \begin{subfigure}[t]{\textwidth}
      \begin{minipage}[t]{0.04\textwidth}
        \vspace{0pt}
        \caption{}\label{fig:e-a_20.0}
      \end{minipage}
      \hfill
      \begin{minipage}[t]{0.95\textwidth}
        \vspace{0pt}
        \includegraphics[width=\textwidth]{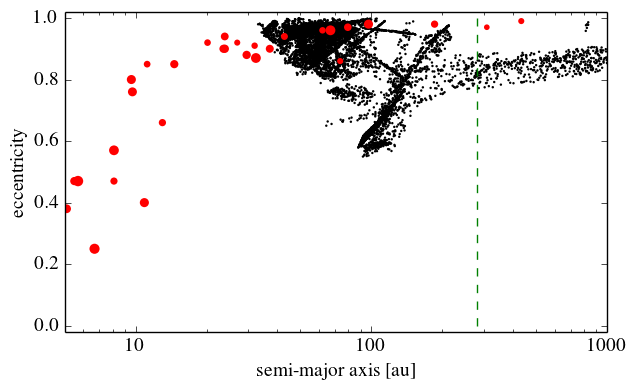}
      \end{minipage}
    \end{subfigure}
  \end{minipage}
  \caption{Distribution of counter-orbiting objects that can be produced by a parabolic, prograde, coplanar fly-by of a perturber
    with mass ratios of $m = 0.1$ ({\bf a}), $m = 1.0$ ({\bf b}), $m = 20.0$ ({\bf c}),
    and pericentre that results in a mass-density drop at $\approx 30$~au (black dots).
    The red circles show the solar system minor bodies from Table~\ref{tab:minor_bodies}.
  }
  \label{fig:e-a}
\end{figure}

Several properties of the solar system indicate that it was severely perturbed by a close fly-by in the past \citep[e.g.][]{2005Icar..173..559M,2010ARA&A..48...47A}.
In addition to the highly eccentric and inclined orbits of TNOs such as Sedna and \mbox{2004 VN$_{112}$} \citep[e.g.][]{2004Natur.432..598K}, for example,
one important property here is the significant decrease in mass (-density) 
by a factor of about 1000 \citep{2003EM&P...92....1M}
beyond Neptune at $\approx 30$~au.
In the context of this paper, the question arises whether such an encounter could also have produced retrograde objects in the solar system.

If the mass drop in the solar system at \mbox{$\approx 30$~au} was produced by a prograde, coplanar, parabolic fly-by,
Eq.~(10) from \citet{2014A&A...565A.130B} can be used to estimate the pericentre distance of the interaction.
The pericentre distance is then given by
\begin{align}
  r_{\mathrm{p,peri}}  = \frac{30~\mathrm{au}}{0.28} {m}^{0.32}.
\end{align}
For a perturber with a mass of $M_{\mathrm{p}} =0.1~M_{\odot}$, this results in a minimum estimate for the pericentre distance of $r_{\mathrm{p,peri}}  \approx 51$~au.
For the mass-ratios of $m = 1.0$ and $20.0$, we obtain minimum pericentre distances of $\approx 107$~au and $\approx 279$~au.

Figure~\ref{fig:e-a} shows the eccentricities and semi-major axes of the counter-orbiting objects produced in a prograde,
coplanar encounter with the above periastron distances for the three mass-ratios.
The distribution of final values in the a-e plane (black dots) was obtained by sampling $10000$ VPPs
with $0.1 < r_{\mathrm{vpp}}  < 10$ for which the final orbit is counter-orbiting the host and the final semi-major axis is $\le 1000$~au.
Initial radii of $> 1000$~au were also rejected.
This large number of values has been sampled to obtain a good coverage of the region of the a-e plane, which may be populated by such objects.
The final eccentricities and semi-major axes were interpolated from our simulation results and scaled by the above pericentre distances.
The vertical green dashed line shows the pericentre distance of the fly-by.

We compare the objects shown in Fig.~\ref{fig:e-a} to the solar system minor bodies with semi-major axes $\gtrsim 5$~au and inclinations $> 140^{\circ}$.
These bodies (see Table~\ref{tab:minor_bodies}) are shown in Fig.~\ref{fig:e-a} with red circles.
The size of a circle indicates the orbital inclination of the respective body: the closer to i=180, the larger the dot.
We do not consider the angles between the angular momentum vectors of the minor bodies and the solar spin axis, which would be the true spin-orbit angles,
but only the orbital inclinations of the bodies relative to the ecliptic.
The maximum difference between the inclination and the spin-orbit angle is $\pm \approx 7^{\circ}$.
Because we compare eccentricities and semi-major axes and not inclinations, however, this only has a minor effect here.
Even the correct spin-orbit angles would only affect the symbol sizes.

For $m = 0.1$, the resulting semi-major axes are $\gtrsim 70$~au.
For $m = 1.0$ and $m = 20.0$, they are $\gtrsim 30$~au.
For all three mass-ratios the eccentricities and semi-major axes of a few solar system minor bodies
have values in a range that might have been produced in such an encounter.
The whole distribution of the solar system bodies does not really fit the distributions produced by the encounters with the shown mass ratios, however.

In this context, two things have to be considered:
First, there is an observational bias on the known outer solar system bodies towards objects with smaller semi-major axes \citep[e.g.][]{2005Icar..173..559M}.
In the future, objects with larger semi-major axes may be found, which might be in agreement with one of the simulated populations.
Second, the eccentricities and semi-major axes as produced by an encounter are only distributed as shown shortly after the encounter.
Processing the perturbed system by secular evolution for $\gtrsim 4$ billion years will alter the orbital elements \citep{2014MNRAS.444.2808P}.
The processed population might fit the current orbital elements of the solar system minor bodies better.
Finally, better fits might be obtainable with inclined fly-bys. This investigation is beyond the scope of this paper, however.

\subsection{Sample application to a planetary system}

\begin{figure}[t!]
  \centering
  \begin{minipage}[t]{\hsize}
    \vspace{0pt}
    \begin{subfigure}[t]{\textwidth}
      \begin{minipage}[t]{0.04\textwidth}
        \vspace{0pt}
          \caption{}\label{fig:sample_system_ecc}
      \end{minipage}
      \hfill
      \begin{minipage}[t]{0.95\textwidth}
        \vspace{0pt}
        \includegraphics[width=\textwidth]{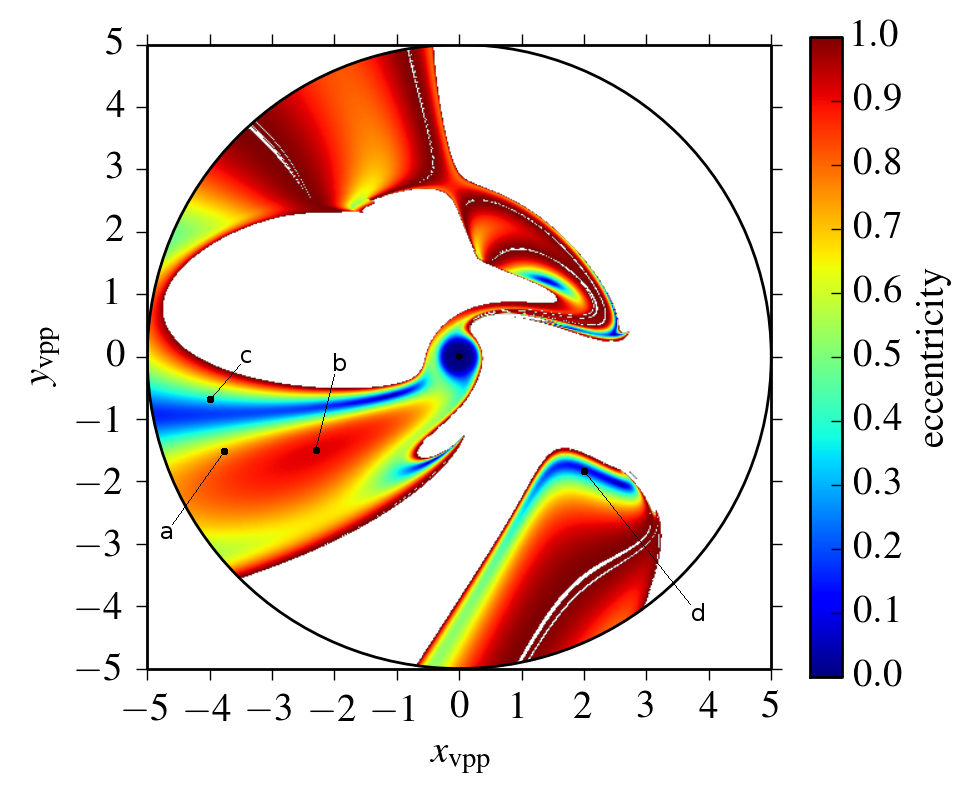}
        \end{minipage}
    \end{subfigure}
  \end{minipage}
  \begin{minipage}[t]{\hsize}
    \vspace{0pt}
    \begin{subfigure}[t]{\textwidth}
      \begin{minipage}[t]{0.04\textwidth}
        \vspace{0pt}
        \caption{}\label{fig:sample_system_semimajor}
      \end{minipage}
      \hfill
      \begin{minipage}[t]{0.95\textwidth}
        \vspace{0pt}
        \includegraphics[width=\textwidth]{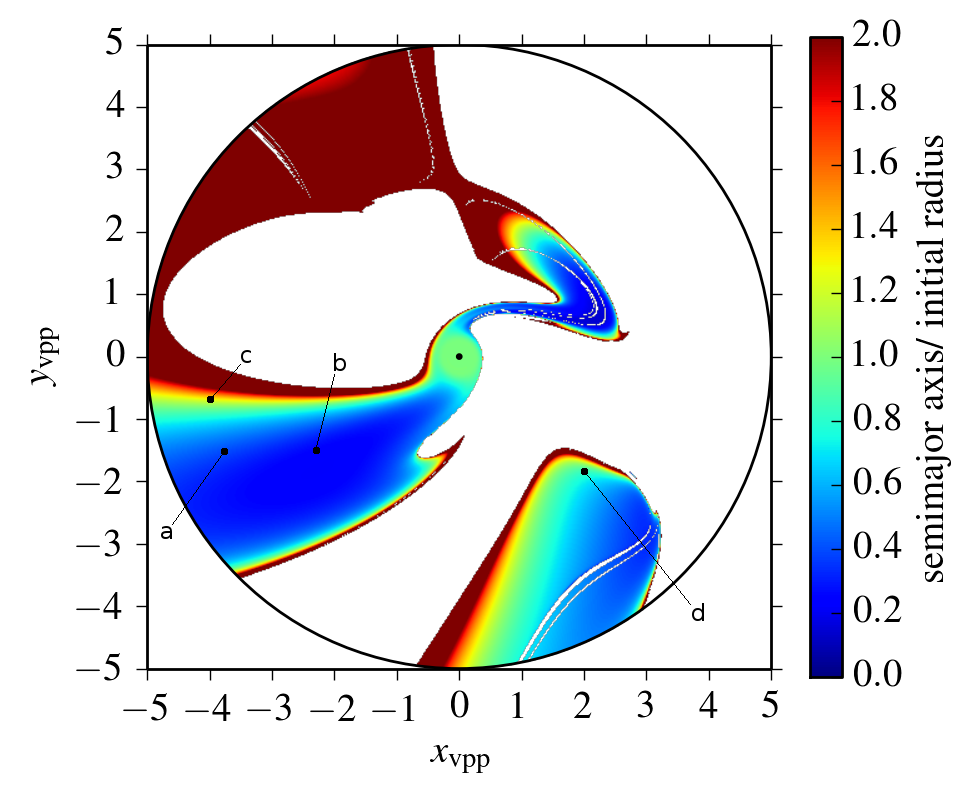}
      \end{minipage}
    \end{subfigure}
  \end{minipage}
  \caption{Position of the sample planets in the VPP space and corresponding final orbital eccentricity ({\bf a}) and semi-major axes ({\bf b}).
    Image~({\bf a}) taken from \cite{2017A&A...599A..91B}, image~({\bf b}) from the Corrigendum \citep{2017A&A...605C...1B}.
    In contrast to the original images, not all particles that are finally bound to host or perturber are shown here,
    but only those that are finally bound to the host.
  }
  \label{fig:sample_system}
\end{figure}

As shown in \citet{2017A&A...599A..91B} and above, the outcome of a perturbation by a fly-by
does not simply depend on the relative initial radii, $r_{\mathrm{init}} $,  of the perturbed matter.
Instead, the full VPPs have to be considered.
For the perturbation of a planetary system by a passing star, this can have counterintuitive consequences
that we demonstrate in the following.

We assume a system with two planets that is perturbed in an equal-mass, parabolic, coplanar, and prograde encounter with a pericentre distance of $r_{\mathrm{p,peri}} $.
One of the planets has an initial orbital radius of $r_{\mathrm{vpp}}  \approx 2.7$, the other of $r_{\mathrm{vpp}}  \approx 4$.
As a result, it is possible that the outer planet ends up on a prograde orbit with $e \approx 0.7$ and $a/r_{\mathrm{vpp}}  \approx 0.4 \Rightarrow a \approx 1.6\, r_{\mathrm{p,peri}} $
and the inner planet on a prograde orbit with $e \approx 0.9$ and $a/r_{\mathrm{vpp}}  \approx 0.3 \Rightarrow a \approx 0.8\, r_{\mathrm{p,peri}} $
(see the points labelled a) and b) in Fig.~\ref{fig:sample_system}).

A much less likely but also possible outcome is that the outer planet ends up on a prograde orbit with $e \approx 0.3$ and $a/r_{\mathrm{vpp}}  \approx 3.2 \Rightarrow a \approx 12.8\, r_{\mathrm{p,peri}} $
and the inner planet on a {\bf retrograde} orbit with $e \approx 0.2$ and $a/r_{\mathrm{vpp}}  \approx 2.2 \Rightarrow a \approx 6\, r_{\mathrm{p,peri}} $
(see the points labelled with c) and d) in Fig.~\ref{fig:sample_system}).
In this case, the final orbital angular momentum of the outer planet would be aligned with the stellar spin axis, and the orbital angular momentum of the inner planet would be flipped by $180^{\circ}$.

This example demonstrates that the outcome of a close fly-by is not necessarily that of the outermost planet being perturbed onto a retrograde orbit,
but it can likewise be an inner planet that ends up in this situation.
However, $r_{\mathrm{vpp}}  = 4$ and $r_{\mathrm{vpp}}  = 2.7$ means that the fly-by occurs with a pericentre distance of $1/4$ of the orbital radius of the outer
and $\approx 2/5$ of the orbital radius of the inner planet.
Considering the estimate of fly-by probabilities depending on the minimum distance in Sect.~\ref{sec:discussion},
this is a possible but extremely unlikely event for a typical planetary system.

\subsection{Perturbation of moons during planet-planet scattering}

Recently, \citet{2018ApJ...852...85H} have proposed that fly-by perturbations could also have an effect
on planetary moons during the planet-planet scattering phase of planetary system formation.
If a planetary system forms with several massive planets, dynamical instabilities can develop
that may lead to close encounters between planets and even to the ejection of one of the involved planets \citep[][]{1996Sci...274..954R}.
When one of the planets has satellites at that time, the orbits of these satellites may be perturbed by the fly-by.
How strong this perturbation is depends here on the mass-ratio and the pericentre distance of the fly-by.
If the satellites initially are on circular Keplerian orbits around the host planet,
the orbit of the fly-by is approximately parabolic,
and the influence of the host star is negligible during the event,
the perturbation process might deviate only slightly from the perturbation of matter around a star by a stellar mass perturber.
In this case, similar results as shown by \citet{2017A&A...599A..91B} for stellar fly-bys can be expected.

Consequently, the results presented in this work would also be applicable as first-order approximation.
This means it can be speculated that through the planet-planet scattering mechanism, retrograde moons are produced as well.
Since in this scenario the probability for nearly coplanar fly-bys is higher than in the case of stellar fly-bys in star clusters,
the probability to obtain retrograde objects is also higher ($\approx 10 \text{--} 20$~\%).
As shown above, the probability to end up on a retrograde orbit would be higher for moons with larger initial orbital radii.
This mechanism could also be an alternative formation scenario to capture for the large retrograde Neptun moon Triton or the large number of small retrograde outer moons of the solar system gas giants.
However, this would require more detailed investigations in the future.

\section{Summary and conclusion}
\label{sec:summary}

We investigated the production of retrograde ($\varphi > 90^{\circ}$) and counter-orbiting ( $\varphi \approx 180^{\circ}$) planets by parabolic fly-bys of stars.
Considering different inclinations between the perturber orbit and the orbital plane of the perturbed matter,
we found the probability to produce retrograde or counter-orbiting matter to be in general very low for low-mass perturbers.
For equal-mass and high-mass perturbers, the probability to produce retrograde objects is about $10\%$,
the production of counter-orbiting objects, that is, the flipping of the orbital angular momentum vector of the planet, is possible in encounters with $i \approx 0^{\circ}$ or $i \approx 180^{\circ}$.
Surprisingly, for these mass-ratios, counter-orbiting objects are more likely to be produced in prograde, coplanar fly-bys ($i = 0^{\circ}$) than in retrograde, coplanar ones ($i = 180^{\circ}$).
We found the highest probability of $\approx 13\%$ for equal-mass perturbers on prograde, coplanar orbits.

We studied this relatively high probability to produce counter-orbiting matter in prograde fly-bys in more detail by investigating the trajectories leading to this outcome in a special reference frame.
For the investigated mass-ratios we found several different mechanisms that may turn initially prograde into retrograde orbits.
The mechanism acting on a certain particle depends sensitively on the virtual pericentre position of the particle.
For the equal-mass case, we found a range of initial orbital radii for which it is more likely for an object to become counter-orbiting the host than to remain on a prograde orbit.

To judge the general significance of the production of retrograde and counter-orbiting objects in fly-bys,
we also investigated the probability of these events as a function of radius,
averaged over the inclination of the perturber orbit.
The probability to produce a retrograde object in a close fly-by is $\approx 10\%$ for mass ratios of $m \gtrsim 1$.
The probability to produce a counter-orbiting object is about $1\%$.

Considering also the range of orbital elements of fly-bys, as typical in young star clusters,
the production of retrograde and counter-orbiting planets is possible, but with a relatively low probability.
For a more detailed statement about the probability to produce retrograde planets in fly-bys in star clusters,
a more sophisticated analysis is necessary.

Finally, we have investigated some practical consequences of our results.
For the counter-orbiting objects in the solar system, we found only a very low probability that their current orbital elements were produced by a prograde, coplanar fly-by of another star.
Especially their current semi-major axes are too small to be caused by an encounter that could also be responsible for the mass drop at the outer edge of the solar system.

We have seen that mostly very close fly-bys lead to retrograde or even counter-rotating orbits.
Such close fly-bys are much more common in very dense clusters that eventually develop into open clusters than in the typical clusters in the solar neighbourhood \citep[][]{2018ApJ...868....1V}.
One prediction of the model we presented here would be that retrograde and counter-orbiting planets should be more common around stars in open clusters than around field stars.

\vspace{1em}
{\tiny {\bf Acknowledgements}:
  We thank the anonymous referee, whose constructive suggestions and comments we highly appreciate.
  This research has made use of data provided by the International Astronomical Union Minor Planet Center.
}

\bibpunct{(}{)}{;}{a}{}{,} 

\bibliographystyle{aa}

\begin{appendix}

\section{More about the rotating reference frame}
\label{sec:more_rotating}

Let the position of the host at time $t$ be $(x_{\mathrm{h}}(t),\,y_{\mathrm{h}}(t))$,
the position of the perturber $(x_{\mathrm{p}}(t),\,y_{\mathrm{p}}(t))$,
and the position of the particle $(x(t),\,y(t))$.
The position of the host in the host-centred rotating reference frame is always (0,\,0)
and the position of the perturber is (1,\,0).
The position of the particle in this system is then given in polar coordinates by
\begin{align}
  \begin{split}
    r_{\mathrm{rot}}(t) &= \frac{r(t)}{r_{\mathrm{p}}(t)}\\
    &= \frac{\sqrt{(x(t)-x_{\mathrm{h}}(t))^2 + (y(t)-y_{\mathrm{h}}(t))^2}}
    {\sqrt{(x_{\mathrm{p}}(t)-x_{\mathrm{h}}(t))^2 + (y_{\mathrm{p}}(t)-y_{\mathrm{h}}(t))^2}}
  \end{split}
\end{align}
and
\begin{align}
  \begin{split}
    \theta_{\mathrm{rot}}(t) &= \theta(t) - \theta_{\mathrm{p}}(t)\\
    &= \mathrm{atan2}(x(t)-x_{\mathrm{h}}(t),\,y(t)-y_{\mathrm{h}}(t)) -\\
    &\quad\,\,\mathrm{atan2}(x_{\mathrm{p}}(t)-x_{\mathrm{h}}(t),\,y_{\mathrm{p}}(t)-y_{\mathrm{h}}(t)).
  \end{split}
\end{align}
The Cartesian coordinates of the particle in the rotating reference frame are given by
\begin{align}
  \chi(t) = r_{\mathrm{rot}}(t) \cos{(\theta_{\mathrm{rot}}(t))}
\end{align}
and
\begin{align}
  \Psi(t) = r_{\mathrm{rot}}(t) \sin{(\theta_{\mathrm{rot}}(t)).}
\end{align}
We note that in contrast to other definitions of the rotating reference frame, our system is host centred
and not centred on the centre of mass.

\section{Interactions ending in retrograde orbits for $m = 1.0$}
\label{sec:r3f_m1}

\begin{figure}[ht!]
  \centering
  \begin{minipage}[t]{\hsize}
    \vspace{0pt}
    \begin{subfigure}[t]{\textwidth}
      \begin{minipage}[t]{0.04\textwidth}
        \vspace{0pt}
          \caption{}\label{fig:C}
      \end{minipage}
      \hfill
      \begin{minipage}[t]{0.95\textwidth}
        \vspace{0pt}
        \centering
        \includegraphics[width=0.8\textwidth]{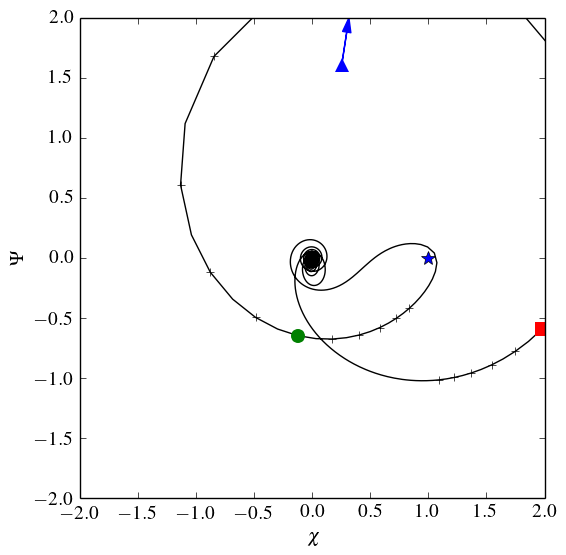}
        \end{minipage}
    \end{subfigure}
  \end{minipage}
  \begin{minipage}[t]{\hsize}
    \vspace{0pt}
    \begin{subfigure}[t]{\textwidth}
      \begin{minipage}[t]{0.04\textwidth}
        \vspace{0pt}
        \caption{}\label{fig:D}
      \end{minipage}
      \hfill
      \begin{minipage}[t]{0.95\textwidth}
        \vspace{0pt}
        \centering
        \includegraphics[width=0.8\textwidth]{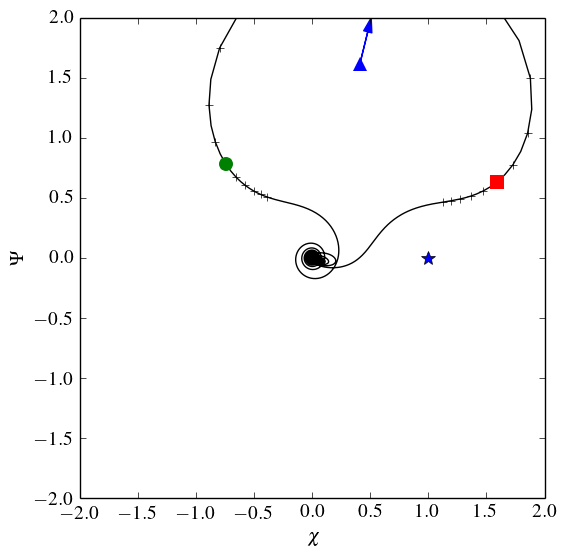}
      \end{minipage}
    \end{subfigure}
  \end{minipage}
  \begin{minipage}[t]{\hsize}
    \vspace{0pt}
    \begin{subfigure}[t]{\textwidth}
      \begin{minipage}[t]{0.04\textwidth}
        \vspace{0pt}
        \caption{}\label{fig:E}
      \end{minipage}
      \hfill
      \begin{minipage}[t]{0.95\textwidth}
        \vspace{0pt}
        \centering
        \includegraphics[width=0.8\textwidth]{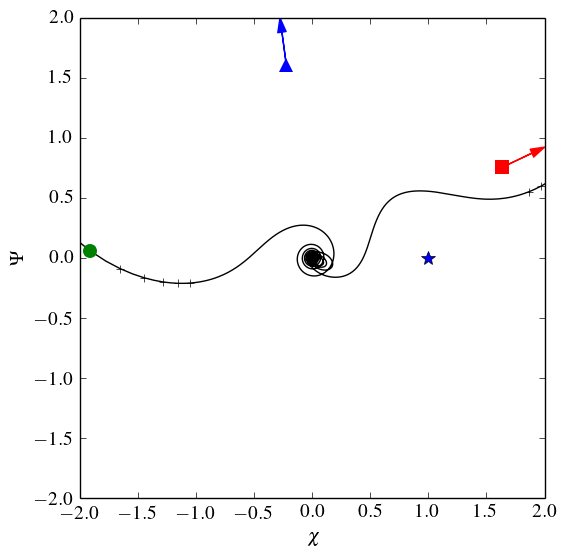}
      \end{minipage}
    \end{subfigure}
  \end{minipage}
  \caption{Interactions leading to retrograde orbits in the R3F for the cases\\
    \mbox{{\bf C (a)}: $i = 0$, $m = 1.0$, $x_{\mathrm{vpp}}  = 1$, $y_{\mathrm{vpp}}  = -3$},\\
    \mbox{{\bf D (b)}: $i = 0$, $m = 1.0$, $x_{\mathrm{vpp}}  = 1$, $y_{\mathrm{vpp}}  = 2$}, and\\
    \mbox{{\bf E (c)}: $i = 0$, $m = 1.0$, $x_{\mathrm{vpp}}  = -1$, $y_{\mathrm{vpp}}  = 3$}.\\
    The green circles, blue triangles, and red squares mark the positions of the particles for $t = -t_{\mathrm{slr}}$, $t= 0$, and $t = t_{\mathrm{slr}}$ respectively.
    The definition of the Cartesian coordinates of the R3F, $\chi$ and $\Psi$,
    can be found in Sect.~\ref{sec:more_rotating}.\\ 
  }
  \label{fig:C_D_E}
\end{figure}

The particles from the red region close to the approaching branch of the perturber orbit undergo a similar interaction with host and perturber as the
particles from the corresponding region in Fig.~\ref{fig:map_0.1} for the mass ratio of $m = 0.1$.
Figure~\ref{fig:C} shows the interaction in the rotating reference frame for a particle with VPPs of $x_{\mathrm{vpp}}  = 1$, $y_{\mathrm{vpp}}  = -3$ (black dot labelled with {\bf C} in Fig.~\ref{fig:map_1}).
The trajectory is very similar to the trajectory in Fig.~\ref{fig:A}.
The particle also has a relatively strong clockwise interaction with the perturber before $t = -t_{\mathrm{slr}}$ and moves retrograde around the host while the perturber passes pericentre.
Instead of passing between host and perturber when the perturber is close to $SLR$, however, the particle passes the perturber outside of the perturber orbit.

The trajectories of particles from the red regions close to the departing branch of the perturber orbit in Fig.~\ref{fig:map_1} (cases {\bf D} and {\bf E})
are similar to the trajectory of case {\bf B}.
Like in case {\bf B}, the particles move on relatively unperturbed orbits with different orbital radii while the perturber passes $-SLR$ and pericentre.
After the perturber passed $SLR$, it passes behind the particle and attracts it just long enough to change its sense of rotation around the host.

\end{appendix}

\end{document}